\documentclass[prx,twocolumn,preprintnumbers,amsmath,amssymb,superscriptaddress]{revtex4-1}
\usepackage{graphicx, epsf,bm,amsmath}

\usepackage{subfigure,color}
\usepackage{amsmath,amssymb}
\usepackage{graphicx}
\usepackage{wasysym}
\usepackage{amsfonts}
\usepackage{bm}
\usepackage{enumerate}
\usepackage{color}
\usepackage[dvipsnames]{xcolor}
\usepackage[resetlabels]{multibib}
\usepackage{epstopdf}
\usepackage{latexsym}
\usepackage[breaklinks,colorlinks = true,linkcolor = red,urlcolor=cyan,citecolor=red]{hyperref}

\newcommand{\br}{{\bf r}}

\newcommand{\bq}{{\bf q}}

\usepackage{braket}
\usepackage{units}

\begin{document}
\title{Disorder in Twisted Bilayer Graphene}
\author{Justin H. Wilson}
\affiliation{Department of Physics and Astronomy, Center for Materials Theory, Rutgers University, Piscataway, NJ 08854 USA}
\author{Yixing Fu}
\affiliation{Department of Physics and Astronomy, Center for Materials Theory, Rutgers University, Piscataway, NJ 08854 USA}
\author{S. Das Sarma}
\affiliation{Condensed Matter Theory Center and Joint Quantum Institute, Department of Physics, University of Maryland, College Park, MD 20742, USA}
\author{J. H. Pixley}
\affiliation{Department of Physics and Astronomy, Center for Materials Theory, Rutgers University, Piscataway, NJ 08854 USA}
\date{\today}

\begin{abstract}
We develop a theory for a qualitatively new type of disorder in condensed matter systems arising from local twist-angle fluctuations in two strongly coupled van der Waals monolayers twisted with respect to each other to create a flat band moir\'e superlattice.
The new paradigm of `twist angle disorder' arises from the currently ongoing intense research activity in the physics of twisted bilayer graphene.
In experimental samples of pristine twisted bilayer graphene, which are nominally free of impurities and defects, the main source of disorder is believed to arise from the unavoidable and uncontrollable non-uniformity of the twist angle across the sample.
To address this new physics of twist-angle disorder, we develop a real-space, microscopic model of twisted bilayer graphene where the angle enters as a free parameter.
In particular, we focus on the size of single-particle energy gaps separating the miniband from the rest of the spectrum, the Van Hove peaks, the renormalized Dirac cone velocity near charge neutrality, and the minibandwidth.
We find that the energy gaps and minibandwidth are strongly affected by disorder while the renormalized velocity remains virtually unchanged.
We discuss the implications of our results for the ongoing experiments on twisted bilayer graphene.
Our theory is readily generalized to future studies of twist angle disorder effects on all electronic properties of moir\'e superlattices created by twisting two coupled van der Waals materials with respect to each other.
\end{abstract}

\maketitle

\section{Introduction}

The ability to isolate and characterize single sheets of graphene~\cite{Geim-2009} has lead to a significant amount of control over van der Waals heterostructures~\cite{Geim-2013}.
This spectacular materials engineering feat has led not only to relatively clean, high mobility graphene samples, but also to the ability to place two coupled sheets of graphene on top of each other with a relative ``twist'' angle between them~\cite{li_observation_2010}.
The introduction of the `twist angle' as a new experimental parameter to tune the electronic properties of `twisted' van der Waals heterostructures has led to a new paradigm in condensed matter systems where one can now study materials properties not only as a function of temperature, carrier density, magnetic field, gate voltage, applied pressure or strain, etc., but also as a function of the twist angle between the two layers controlling the electronic band structure in a radical manner, which is a completely new tool in the laboratory.  This new tool of a variable twist angle is revolutionizing condensed matter physics, leading to many new discoveries in twisted bilayer graphene every month.
Since the initial experiments on twisted bilayer graphene establishing the fabrication technique~\cite{Luican-2011}, recent experiments have shown that the system can support purportedly correlated insulating states and superconductivity~\cite{CaoJarillo2018a,CaoJarillo2018b,Yankowitz-2019,Lu-2019}.
While these bilayers are clean and relatively disorder free, the twist angle can vary across samples, leading to a new source of disorder.
Thus, even if the two starting monolayers are completely clean (i.e. no impurities or defects), the very fact of creating the twisted bilayer system introduces an inherent (and a new type of) disorder by virtue of local fluctuations in the twist angle throughout the macroscopic sample.  This `twist angle disorder', which has no analogy in usual condensed matter systems and has never before been studied in the literature, is thought to be the main disorder controlling the quality of the currently available twisted graphene systems.

In single-layer graphene, the most dominant effect of disorder near the Dirac point has been attributed to charge disorder (arising from unintentional quenched random charged impurities in the system) inducing ``puddles'' of unequal charge density that locally dope the Dirac cones~\cite{DasSarma-2011}.
This issue has recently been circumvented by using an all van der Waals device geometry, and the absence of any significant charge inhomogeneities in such ultra-clean samples has enabled the observation of exotic many-body states~\cite{Zibrov-2017} akin to what has been seen in clean suspended graphene~\cite{Du-2009}.
As a result, the current graphene sample quality is rather remarkable and for most practical purposes, both charge inhomogeneities as well as any extrinsic disorder due to vacancies or defects has been greatly suppressed, if not almost eliminated except perhaps for experiments using very low ($<\unit[10^{10}]{cm^{-2}}$) carrier densities.

With these capabilities, very clean samples of twisted bilayer graphene (TBG) near the magic-angle (where the nominal band structure becomes completely flat suppressing the Dirac velocity to zero) have recently been observed to develop insulating states at integer filling fractions of the moir\'e miniband near the Dirac points~\cite{CaoJarillo2018a,Yankowitz-2019}.
Upon gating (i.e.\ doping) away from the insulating phases, nearby superconducting phases have been observed~\cite{CaoJarillo2018b,Yankowitz-2019,Lu-2019}.
To achieve an accurate choice (to within $\sim 0.1^{\circ}$) and rather small value  of the twist angle ($\sim 1^\circ$), the ``tear and stack'' mechanical approach places two sheets of graphene on top of each other with a great deal of precision in the twist angle~\cite{Kim-2016}.
Only after such a mechanical procedure of creating the twisted bilayer sample with a carefully chosen twist angle, the sample is transferred to the cryostat for electrical measurements.
To study the electronic properties as a function of the twist angle, the whole procedure has to be repeated for a different sample with a different twist angle.
In practice however, this procedure does not produce a single twist angle across the entire sample:  Scanning tunneling microscopy has observed different twist angles across separate regions of the sample~\cite{li_observation_2010,brihuega_unraveling_2012,wong_local_2015,kerelsky_magic_2018,choi_imaging_2019,jiang_charge-order_2019,xie_spectroscopic_2019}.
Moreover, signatures of the nonuniformity of the twist angle have also been observed in conductance measurements that have a strong dependence on where the leads are placed on the device~\cite{Yankowitz-2019}.
In addition, two different samples with nominally identical twist angles typically manifest quite different electronic properties in transport and STM measurements, again reflecting that some inherent variations in the twist angle invariably exist in the system.
Thus, in any given high-quality (i.e.\ low extrinsic impurity and defect concentration) sample the main source of disorder comes in the form of a \emph{varying} twist angle across the sample.
The nature of this variation is not unique: some samples have hard domain walls separating regimes with different twist angles, whereas some samples have a much smoother change in the twist across the sample \cite{Priv_Comm_Eva}.
Currently, the qualitative effect of such forms of disorder on the single-particle spectrum near the magic angle is unknown and to the best of our knowledge, there has not been any attempt to describe this in a precise fashion.
Twist angle disorder is a radically new type of intrinsic disorder in condensed matter systems whose study is, quite apart from its singular importance in determining the twisted graphene bilayer properties, of fundamental conceptual significance.

The numerical study of twist-angle disorder is difficult with the current models available in the literature.
First, the usual continuum model is built as a hexagonal lattice in momentum-space \cite{lopesdossantos_graphene_2007,BistritzerMacDonald2011} where disorder enters the Hamiltonian in a highly non-local way.
Second, current real-space models rely on both a uniform and commensurate twist angle \cite{mele_commensuration_2010,moon_energy_2012}.
To circumvent this problem, we build a new real space model where the twist is built directly into the interlayer hopping in such a way that it can be continuously tuned, and can vary spatially while the model remains \emph{local} in real space.
The model exactly reproduces the continuum model as written by Bistritzer and MacDonald~\cite{BistritzerMacDonald2011} near the $K$ and $K'$ Dirac points in the Brillouin zone.
The version of this model presented here preserves $C_2 T$ symmetry (i.e. the combined operation of a 180$^{\circ}$ rotation and a time reversal operation) and hence preserves the Dirac nodes.
Further, it qualitatively preserves the spatial structure of AA and AB tunneling; however, it explicitly breaks $C_3$ symmetry.
While there is no obstruction to building the model with $C_3$ symmetry (a version of which will appear in Ref.~\cite{Fu-2018}), real experiments introduce strain which also explicitly breaks $C_3$ \cite{bi_designing_2019}, so we do not require this of our real space model.
So, in some sense, our disorder model incorporates both the twist-angle disorder (in a controlled manner) and strain effects (in an uncontrolled manner through the explicit breaking of $C_3$ symmetry).

\begin{figure}
\includegraphics[width=\columnwidth]{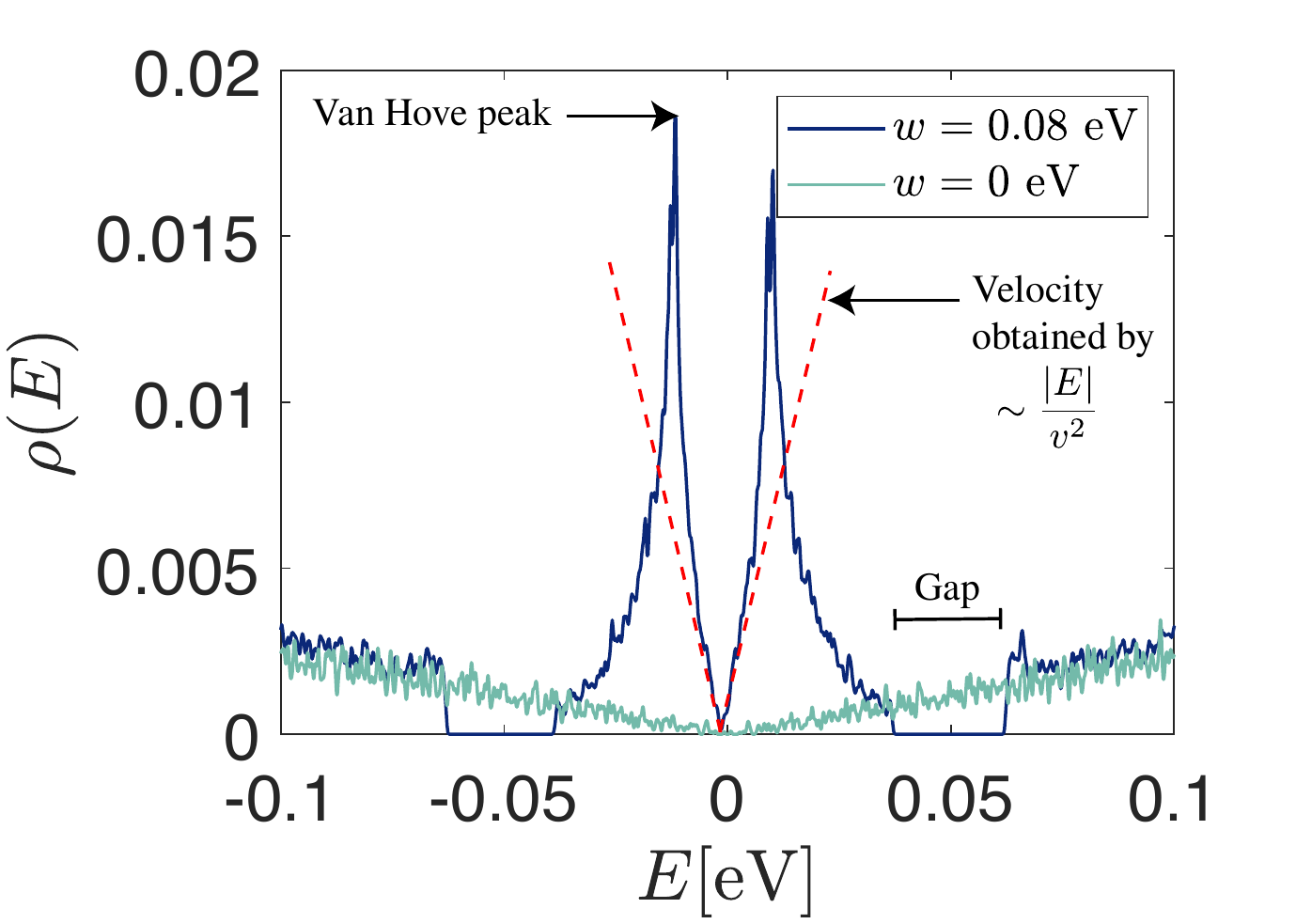}
\caption{(color online) The density of states $\rho(E)$ as a function of energy $E$ for the  lattice model of twisted bilayer graphene  at a twist angle $\theta = 1.05^{\circ}$, a linear system size $L=569$, a kernel polynomial method~\cite{Weisse-2006} expansion order $N_C=2^{17}$, and a weak breaking in the interlayer tunneling between AA and AB sites ($w_0/w_1 = 0.75$, $w_1=w$ where $w_0$ ($w_1$) is the strength of AA and BB (AB and BA) tunneling), which captures lattice relaxation effects~\cite{Nam-2017,Carr-2019} and it opens a hard gap on both sides of the semimetal miniband.
We note that at small angles, a single parameter controls the physics: $w/[2 v_F k_D \sin(\theta/2)]$, so lowering the angle is equivalent to increasing $w_1$. Therefore, one can read the plots of smaller $w_1$ as at an angle larger than $1.05^{\circ}$.
This density of states has a number of features relevant to the physics: Van Hove peaks, gaps, and the velocity (as determined by the scaling of the density of states).
Dark (light) blue lines give the calculated density of states for finite (zero) values of the parameter $w$ as shown in the inset of the figure.}
\label{fig:cleanDOS}
\end{figure}

Once we have a viable real-space model, we are able to study the  effects of disorder  on twisted bilayer graphene at the single-particle level.
Not only does twisting two sheets of graphene create flat bands near the magic-angle ($\sim1$-$1.1^\circ$), it also induces gaps that separate the miniband, which has Van Hove singularities in the density of states~\cite{Li-2010,BistritzerMacDonald2011,DosSantosNeto2012}, from the rest of the spectra as  seen in Fig.~\ref{fig:cleanDOS}.
These miniband insulating gaps arising from the single-particle band structure of the twisted system are simply the moir\'e superlattice band gaps due to the tunneling between the two graphene bands in the combined bilayer heterostructure.
We are interested in how all of these single-particle, superlattice, miniband features are affected or even destroyed due to randomness in the twist angle.
Recently, it has been demonstrated that many of these features can be captured using much simpler models with Dirac points perturbed by a quasiperiodic potential that mimics the twist~\cite{PixleyGopalakrishnan2018, Fu-2018}.
These effective models are rather natural as most twist angles are not commensurate, and hence, a quasiperiodic incommensurate background potential should have effects very similar to the moir\'e potential induced by the twist angle.
In fact, twisted bilayer graphene at a large twist angle ($\sim 30^{\circ}$) has recently been used to form quasicrystals~\cite{Yao-2018,Ahn-2018}, and renormalized but stable low-energy Dirac excitations have been observed~\cite{Ahn-2018}, supporting the idea of an incommensurate quasiperiodic potential mimicking the twist-angle moir\'e superlattice.
These simpler quasiperiodic models exhibit a similar magic-angle condition where the velocity of the Dirac cone vanishes continuously.
In addition, the formation of minibands  with large gaps and a strongly renormalized  velocity that can be seen to clearly vanish without having to resort to very large system sizes as in the case of twisted bilayer graphene.
Therefore, we supplement our calculations on twisted bilayer graphene with similar disorder calculations on a quasiperiodic ``toy'' model
to  determine how our choice to model twist disorder impacts our results (see Appendix~\ref{appendix:SOC}).
The two models produce similar results on disorder effects.

We focus on various features of the low energy density of states and the miniband structure to determine how the single-particle spectrum is modified as a result of randomness in the twist angle.
We demonstrate that disorder smooths the non-analyticities in the density of states, fills in the band-gaps, broadens the minibandwidth, and smears out the Van Hove peaks.
We compare this with the size of the gap isolating the low-energy miniband, the renormalized Dirac velocity, and the size of the minibandwidth. Surprisingly, we find that the Dirac cone velocity is remarkably robust to twist disorder, whereas other miniband characteristics are systematically broadened.
The essential complete protection of the miniband Dirac velocity (at low energy, where the Dirac cone approximation holds) in the twisted bilayer graphene (TBG) against the twist-angle disorder is a rather unexpected finding of our nonperturbative calculations, particularly since all other aspects of the miniband electronic structure are strongly affected by the twist-angle randomness.

This paper is organized as follows: In Sec.~\ref{sec:model} we build an approximate lattice model for twisted bilayer graphene and use it to introduce  real-space disorder in the twist angle.
In Sec.~\ref{sec:results} we discuss the results of the numerical calculations, and in Sec.~\ref{sec:discussion}, we discuss our approximations and the implications of these results for ongoing experiments.
Finally, we conclude in Sec.~\ref{sec:conclusion} with a summary of our results.
Throughout, we take the lattice spacing between neighboring carbon atoms to be unity, which serves as our unit of length. In Appendix~\ref{appendix:SOC} we analyze a simpler model with similar magic-angle phenomena using a deterministic quasiperiodic potential and in Appendix~\ref{appendix:pt} we provide details on the perturbation theory of the twisted bilayer graphene lattice model.

\begin{figure*}
  \centering
  \includegraphics[width=2\columnwidth]{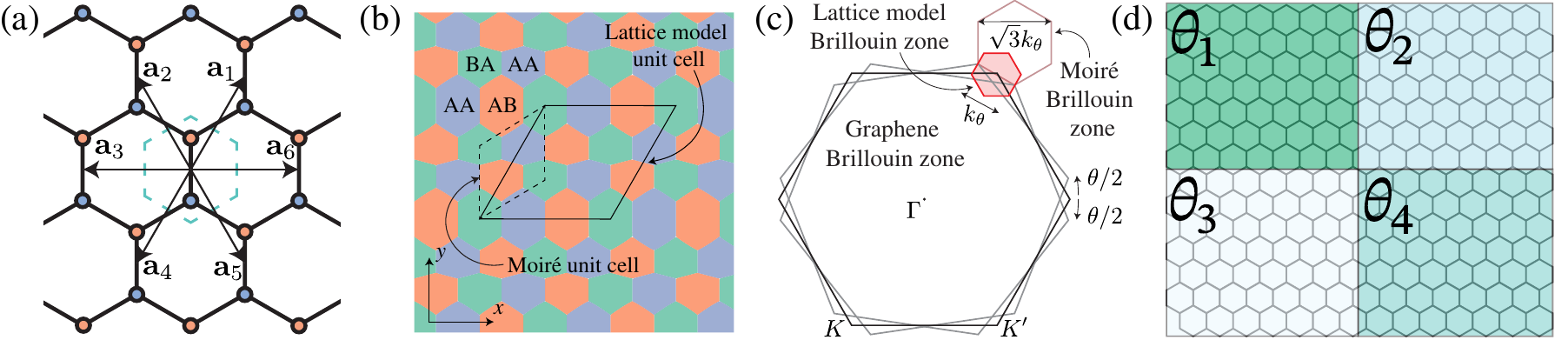}
\caption{(color online) (a) A schematic of graphene and the notation we use for our model. The A (B) sublattice is represented by the blue (orange) lattice sites.
The unit cell for the triangular lattice is shown by the dashed central hexagon.
The lattice vectors are $\mathbf a_1 = (\sqrt{3}/2,3/2)$ and $\mathbf a_2=(-\sqrt{3}/2,3/2)$, and we further define $\mathbf a_3 \equiv \mathbf a_2 - \mathbf a_1$, $\mathbf a_4 \equiv -\mathbf a_1$, $\mathbf a_5 \equiv - \mathbf a_2$, as well as $\mathbf a_6 \equiv - \mathbf a_3$.
(b) A course-grained view of the tunneling between the layers calculated from $\mathcal T_0$ and $\mathcal T_1$ in Eq.~\eqref{eq:Hopping-Ts} which defines the energy parameters $w_0$ and $w_1$; the color represents whether AA, AB, or BA hopping is dominant based on the chance for an electron on a site in layer 1 to hop onto sublattice A or B on layer 2, given by $P_{X}(\mathbf r) = |[\mathcal T_0(\mathbf r)]_{X}|^2 + 6 |[\mathcal T_1(\mathbf r)]_{X}|^2$.
Note that $C_3$ is broken and the moir\'e unit cell is larger than in real TBG.
Both of these effects are relatively small. (c) Complementary to the real space picture, in momentum space the lattice Brillioun zone is effectively downfolded by a factor of three from the moir\'e Brillioun zone after unrotating the two graphene layers; this introduces small gaps in the band structure at these points. (d) In our model, the effect of the twist is entirely contained within interlayer coupling, so we model disorder by changing the continuous twist parameter $\theta$ within different regions of space. In this common example, we break up the system into four equal regions and pick a value of  $\theta_j$ that are drawn from the box distribution $[(1-W_R/2)\theta,(1+W_R/2)\theta]$ with $\theta=1.05^{\circ}$.
}
\label{fig:schematic_all}
\end{figure*}

\section{Model and Approach}\label{sec:model}

The  model (see Fig.~\ref{fig:schematic_all}) we primarily focus on is a lattice model that is an approximation of twisted bilayer graphene which captures the low energy limit of the continuum model~\cite{lopesdossantos_graphene_2007,BistritzerMacDonald2011}.
However, this particular ultra-violet (UV) completion of the continuum model does not respect the underlying $C_3$ symmetry of the microscopic lattice.
As a result, the velocity does not strictly vanish at the magic angle but becomes very small due to the Dirac points not being pinned to high-symmetry points in the Brillouin zone and acquires an angular dependence relative to each Dirac point in momentum space (we show this explicitly using perturbation theory in Appendix~\ref{appendix:pt}).
However, the band structure that results is still qualitatively similar, and so we expect that effects arising from this approximation are not relevant to understand the qualitative effects of disorder.
In any case, it is unclear that a strict magic angle with vanishing velocity can ever be achieved in any laboratory samples, so our approximation of a finite, but very small, velocity should not be a practical problem in any sense.

\begin{figure*}
\centering
\includegraphics[width=2\columnwidth]{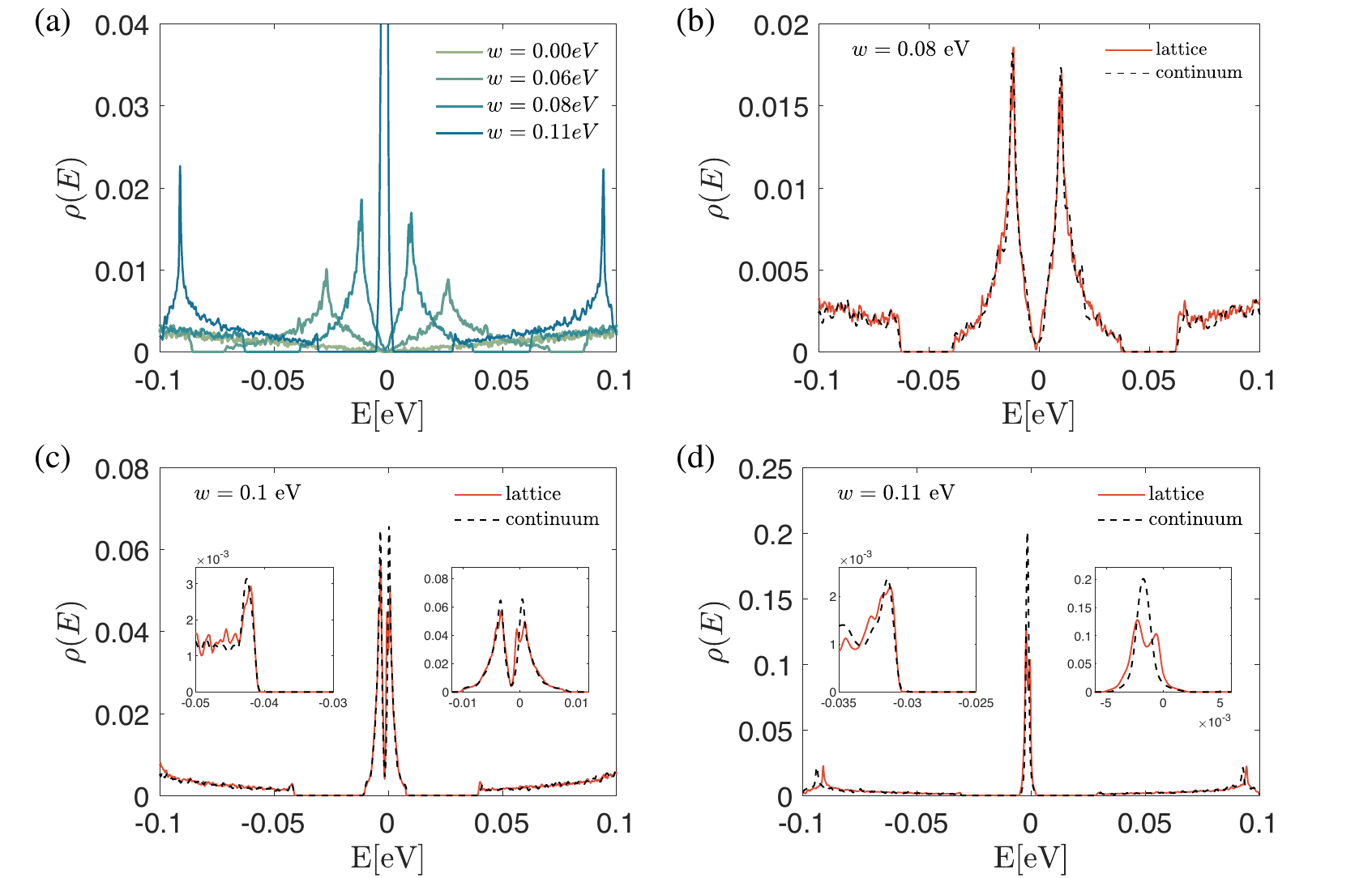}
\caption{(color online) (a) The calculated density of states $\rho(E)$ for TBG without disorder as a function of energy $E$ 
for various interlayer tunneling strengths $w=w_1$ (keeping $w_0/w_1=0.75$ where $w_0$ ($w_1$) denote the strength of AA and BB (AB and BA) tunneling)  at a low twist angle of $\theta= 1.05^{\circ}$ close to the magic angle, a system size of $L=569$ and a KPM expansion order of $N_C=2^{17}$ in the lattice model.
The calculated minibandwidth in the magic angle regime $w=\unit[110]{meV}$ is consistent with other studies of the continuum model and the KPM numerical resolution limits to the extent we can access the low-energy regime near charge neutrality.
(b,c,d) Comparisons between our lattice model and the continuum theory near $E=0$ and $\theta=1.05^{\circ}$ for $w=80$, $w=100$ and $w=\unit[110]{meV}$ respectively, we find remarkable agreement.
The insets show the details of the miniband.
At $\theta=1.05^\circ$ and $w=\unit[100]{meV}$, inset of (c), we see a splitting of the Van Hove peaks that is missing from the continuum model associated with additional zone folding in this model.
This is seen clearly in the right inset; the left inset shows how the gap of the lattice model here and in the continuum model also match rather well.
In (d) at the magic angle $\theta=1.05^\circ$ and $w=\unit[110]{meV}$, we see that the Van Hove peaks never clearly merge as they do in the continuum model. Again, this is clearly seen in the right inset.
The continuum model data here includes 338 bands and has $N_C = 2^{13}$ or $2^{14}$ whereas the lattice model has $L=569$ and $N_C = 2^{17}$.
Overall, the agreement with the continuum TBG model is quite excellent.
}
\label{fig:cleanDOS_TBG}
\end{figure*}

To motivate the model, consider the continuum model written as in Ref.~\onlinecite{BistritzerMacDonald2011} around the $K$ point
\begin{equation}
  H_K =  \begin{pmatrix}
     h_{\mathbf k,\theta/2} &  T(\mathbf r) \\
     T^\dagger(\mathbf r)  &  h_{\mathbf k,-\theta}
\end{pmatrix},
\end{equation}
where $h_{\mathbf k,\theta} = \tfrac{3t}2\mathbf{k} \cdot  e^{-i\theta\sigma_z/2}\bm\sigma^*e^{i\theta\sigma_z/2}$ , $T(\br) = \sum_{j=1}^3 e^{-i (\bq_j\cdot \br+\phi_j)}T_j$ and $T_j = w_0 + w_1(\sigma^+ e^{2\pi i (j-1)/3} + \sigma^- e^{-2\pi i (j-1)/3} )$.
We can ``unrotate'' this Hamiltonian by considering the $\mathbf k$ vectors to be the same and applying a unitary in pseudospin space (using the properties of Dirac cones, one can replace the full angular momentum operator $L_z$ with $\sigma_z/2$)
\begin{equation}
  H_K' =  \begin{pmatrix}
     h_{\mathbf k,0} & e^{i\theta \sigma_z/4} T(\mathbf r) e^{i\theta \sigma_z/4} \\
     e^{-i\theta \sigma_z/4} T^\dagger(\mathbf r) e^{-i\theta \sigma_z/4} &  h_{\mathbf k,0}
\end{pmatrix}.
\end{equation}
We can do a similar operation to the $K'$ point.
The interlayer tunnelings at the $K$ and $K'$ points differ, so it is important to have a function interpolate between the two while preserving symmetries $C_2$ and time-reversal and staying as local as possible.
This can be done, and we can replace the Dirac cone $h_{\mathbf k,0}$ with the Hamiltonian for graphene which in real space and second quantized notation is
\begin{equation}
H_{0}  = \sum_{\br,\ell} t [c^\dagger_{\br,\ell} \sigma_x c_{\br, \ell} +
\sum_{j=1}^2 (c^\dagger_{\br + \mathbf a_j,\ell} \sigma^+ c_{\br, \ell} + \mathrm{h.c.})] \label{eq:freeH}
\end{equation}
where  $t = \unit[2.8]{eV}$ \cite{castro_neto_electronic_2009}, $\br$ labels points on the triangular lattice, $c_{\br,\ell} = (c_{\br,A,\ell}, c_{\br, B,\ell})^T$ is a vector of annihilation operators at triangular lattice site $\br$ and layer $\ell=1,2$ whose first and second components represent the A and B sublattices, respectively.
The lattice vectors $\mathbf a_1$ and $\mathbf a_2$ are shown in Fig.~\ref{fig:schematic_all}(a) where the lattice site $\br$ is the central hexagon. 
The tunneling between layers in real space then becomes
\begin{multline}
  H_{\mathrm{TBG}} = H_0 + \sum_{\br} \left[c_{\br,2}^\dagger \mathcal T_0(\br) c_{\br,1} + \mathrm{h.c.}\right] \\
  + \sum_{\br} \sum_{n=1}^6 \left[(-1)^n c_{\br + \mathbf a_n,2}^\dagger \mathcal T_1(\br + \tfrac12\mathbf a_n)c_{\br,1}+ \mathrm{h.c.}\right]. \label{eq:tbg_ham}
\end{multline}
The second line of of Eq.~\eqref{eq:tbg_ham} represents interlayer hopping to the nearest neighbors on the triangular lattice, summed over all $\mathbf a_n$, as depicted in Fig.~\ref{fig:schematic_all} (a), and the interlayer hopping matrices are given by
\begin{widetext}
\begin{equation}
  \begin{aligned}
  \mathcal T_0(\mathbf r) & = \sum_{j=1}^3\begin{pmatrix}
  w_0 \cos(\mathbf q_j \cdot \mathbf r + \phi_j - \theta/2) & w_1 \cos(\mathbf q_j \cdot \mathbf r - \tfrac{2\pi(j-1)}{3} +\phi_j) \\
  w_1 \cos(\mathbf q_j \cdot \mathbf r + \tfrac{2\pi(j-1)}{3} +\phi_j) &
  w_0 \cos(\mathbf q_j \cdot \mathbf r + \phi_j + \theta/2)
\end{pmatrix} \\
  \mathcal T_1(\mathbf r) & = \frac1{3\sqrt{3}} \sum_{j=1}^3\begin{pmatrix}
  w_0 \sin(\mathbf q_j \cdot \mathbf r + \phi_j - \theta/2) & w_1 \sin(\mathbf q_j \cdot \mathbf r - \tfrac{2\pi(j-1)}{3} +\phi_j) \\
  w_1 \sin(\mathbf q_j \cdot \mathbf r + \tfrac{2\pi(j-1)}{3} +\phi_j) &
  w_0 \sin(\mathbf q_j \cdot \mathbf r + \phi_j + \theta/2)
\end{pmatrix},
\end{aligned} \label{eq:Hopping-Ts}
\end{equation}
\end{widetext}
where $w_0$ represents AA tunneling, $w_1$ is the AB tunneling (commonly, if we refer to $w$, we are referring to $w_1$ and a fixed $w_0/w_1$ ratio), $\theta$ is the twist angle, $\phi_j$ are random phases which sum to zero and represent the center of rotation
\footnote{The phases always modify terms like $\mathbf q_j \cdot \mathbf r + \phi_j$, and in order to see how they represent a center of rotation, consider an $\mathbf q_j \cdot( \mathbf r - \mathbf r_0)$, then $\phi_j = - \mathbf q_j \cdot \mathbf r $ and $\sum_j \phi_j = 0$. In fact, for any $\phi_{1,2}$ we can determine an $\mathbf r_0$ that creates it.}
, $q_1 = k_\theta(0,-1)$, $q_2 = k_\theta(\sqrt{3}/2,1/2)$, and $q_3 = k_\theta(-\sqrt{3}/2,1/2)$.
The value of the twisted wavevector $k_\theta$ is given by $k_\theta = 2k_D \sin(\theta/2)$ where $k_D = 4\pi/(3\sqrt{3})$. The effect of varying $w$ for a fixed twist angle $\theta=1.05^{\circ}$ is shown in Fig.~\ref{fig:cleanDOS_TBG}(a), which demonstrates the formation of a semimetal miniband and shrinking minibandwidth. 
We note that other parameter sets for the tight binding parameters are available~\cite{Lin-2018} but do not affect any of the qualitative results presented here.

If we go to the crystal momentum basis and expand about the $K$ point (with similar results at $K'$), we indeed obtain the continuum model \cite{BistritzerMacDonald2011} up to a unitary transformation as our construction dictated.
Furthermore, if we compare the low-energy continuum model to the actual lattice model itself, we find remarkable agreement in the calculated density of states [defined in Eq.~\eqref{eq:DOS}] as shown in Fig.~\ref{fig:cleanDOS_TBG}(b,c,d) for three representative sets of parameters.

Some comments are in order.
First, while it reproduces the continuum model at the $K$ (and $K'$) point, this particular UV-completion explicitly breaks the $C_3$ symmetry present in the original model (this symmetry is just weakly broken near the $K$ and $K'$ points).
To see this explicitly, we can consider the pattern of AA, AB, and BA tunnelings our model exhibits.
This can be entirely determined by the form of $\mathcal T_0$ and $\mathcal T_1$ in Eq.~\eqref{eq:Hopping-Ts}: If an electron is on an $A$ site on layer 1 and wants to hop to an $A$ site on layer 2, then the sum of the squares of the hoppings give that $P_{AA}(\mathbf r) = |[\mathcal T_0(\mathbf r)]_{AA}|^2 + 6|[\mathcal T_1(\mathbf r)]_{AA}|^2$ (and similarly for $P_{AB}(\mathbf r)$ and $P_{BA}(\mathbf r)$).
Comparing which term [$P_{AA}(\mathbf r)$, $P_{AB}(\mathbf r)$, or $P_{BA}(\mathbf r)$] is largest gives us Fig.~\ref{fig:schematic_all}(b) where we can explicitly see how $C_3$ is broken for this model.
As a result of this symmetry breaking, the Dirac points are not pinned to the high-symmetry points and are free to move around the Brillouin zone, yet since the model preserves the $C_2T$ symmetry they do not gap out.
Numerically, we find that the Van Hove peaks never fully merge [Fig.~\ref{fig:cleanDOS_TBG}(d)] unlike the continuum model, and further, perturbation theory can be used at second order in the interlayer tunneling strength to see that the velocity never fully vanishes either (see Appendix~\ref{appendix:pt}).
For ideal theoretical calculations this might pose a problem, however for our present study, disorder already breaks this $C_3$ symmetry.
Furthermore, in the experimental samples strain from the substrate explicitly breaks the $C_3$ symmetry \cite{bi_designing_2019}, which is a natural single-particle source for a nonvanishing velocity and further justifies the use of this model.
Thus, the weak breaking of $C_3$ symmetry in our model is not a problem at all in understanding the physics of real twisted bilayer graphene systems.
Second, while the model still has the spatial structure of AA, AB, and BA tunneling, it is slightly distorted, as seen in Fig.~\ref{fig:schematic_all}(b).
Consequently, the usual TBG moir\'e unit cell is larger than the unit cell considered in this model.
In fact, the mini-Brillioun zone is folded more than in actual TBG as seen in Fig.~\ref{fig:schematic_all}(c) (it is smaller by a factor of 3); the process of ``unrotating'' the two-layers puts the $K$ points of each individual layer on top of each other in momentum space effectively downfolding the moir\'e Brillouin zone.
It is then necessary to determine if this downfolding opens up any gaps, and while it does, these are small indirect gaps in the mini-Brillioun zone of TBG as seen near the Van Hove peaks in Fig.~\ref{fig:cleanDOS_TBG}(c)(inset).
Last, for the value of the clean twist angle we focus on here $\theta=1.05^{\circ}$ we can emulate the effects of strain and lattice relaxation, similar to Ref.~\cite{Tarnopolsky-2019}, by setting the ratio of AA to AB/BA tunneling to $w_0/w_1 = 0.75$ and $w_1 = w$ based on relaxed band structure calculations~\cite{Nam-2017,Carr-2019}. This acknowledges the empirical fact that Bernal-stacked graphene is the energetically favored stacking arrangement in untwisted bilayer graphene. While varying the twist angle changes the ratio $w_0/w_1$ (as in Refs.\cite{Nam-2017,Carr-2019}), for simplicity we fix this ratio to take that of the clean twist angle ($w_0/w_1=0.75$) throughout.

We compare the lattice model with the continuum model in  Fig.~\ref{fig:cleanDOS_TBG}(b,c,d). We find good agreement between the two models over a rather broad energy range even beyond the low-energy miniband.
In particular, we find that the TBG gap and Dirac velocity are well-produced by the lattice model, see the insets in Fig.~\ref{fig:cleanDOS_TBG}(c,d) . However, a direct comparison at the magic angle condition ($w=0.11$ eV and $\theta=1.05^{\circ}$) reveals that the mini-bandwidth is slightly overestimated within the lattice model.
We further notice that beyond the ``magic-angle'' (i.e.\ smaller angle $\theta$ at fixed $w_1$ or larger interlayer tunneling $w_1$ at fixed $\theta=1.05^{\circ}$), the lack of symmetries leads to disagreement with the continuum model (not shown).
As a result, we restrict ourselves to the regime where the dimensionless parameter $w_1/(k_\theta v_F)$ is below or at  the ``magic'' value where the discrepancy between the continuum model and the effective lattice model is minimized.
Here, we achieve this by focusing on fixing the clean twist angle to $\theta=1.05^{\circ}$ and limit the interlayer tunneling to $w \le 0.11 eV$.
In this regime,
our model captures the TBG electronic structure very well and should be a quantitatively reliable model.
This is also the regime of current experimental interest.

\begin{figure*}
\includegraphics[width=2\columnwidth]{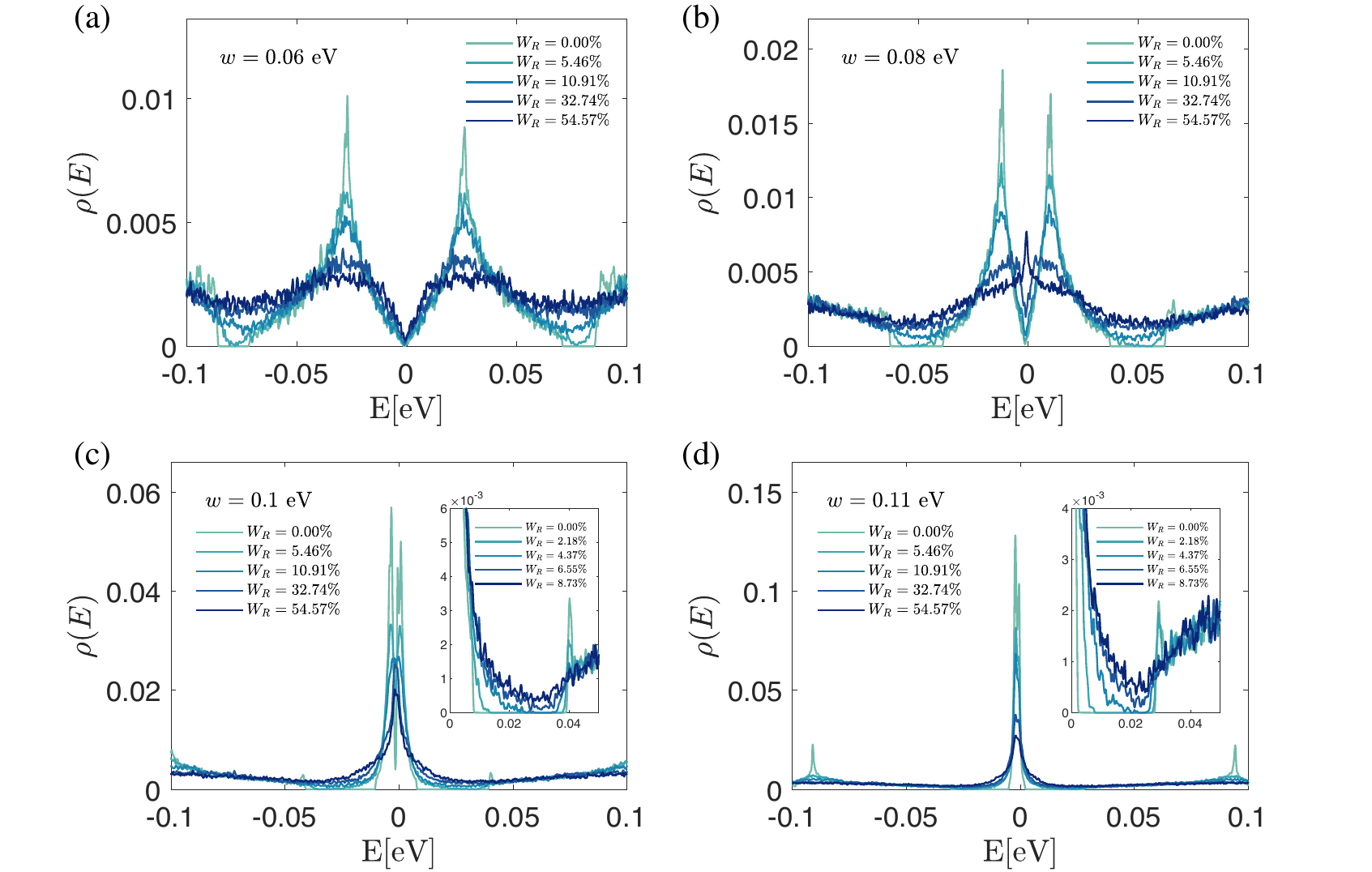}
\caption{The effects of twist disorder on the low energy density of states.
The density of states $\rho(E)$ as a function of energy $E$ for a clean twist angle $\theta=1.05^{\circ}$, linear system size $L=569$, and a KPM expansion order of $N_C=2^{17}$ starting in the semimetal regime of the the TBG model (Top) as well as in the magic-angle regime (Bottom), for different twist-disorder strengths $W_R$ (that characterizes the width of a box distribution $[(1-W_R/2)\theta,(1+W_R/2)\theta]$ with $\theta=1.05^{\circ}$ from which we sample the random twist angle in each patch).
In each case the randomness smoothly fills in the gap while also smearing out the Van Hove peaks.
The insets in the bottom two figures is a zoom in of the band gap that clearly fills in with increasing disorder.
}
\label{fig:DOSE_tbg}
\end{figure*}

In the following, we model the effect of a non-uniform twist by breaking the system up into four equal sections, each having their own twist angle $\theta$ with sharp domains between them, as depicted in Fig.~\ref{fig:schematic_all}(d).
We first choose random phases $\phi_j$ in the interlayer coupling (this reflects different centers of origin for the twist). In what follows, we take a uniform random phase in the TBG calculations as this seems to be the most physically sensible starting point provided the twist angle is not sufficiently small, which would induce significant lattice relaxation~\cite{Nam-2017,Lin2-2018,Carr-2019}.
The $\theta$ in each patch is sampled from a box distribution around a central value $\theta\in [(1 - W_R/2)\theta_0, (1+W_R/2)\theta_0]$ where we express $W_R$ as a percentage and  we fix $\theta_0 = 1.05^{\circ}$.
For twist angles that are small and near the ``magic-angle,'' the moir\'e unit cell includes roughly 10,000 atoms in each layer.
Numerically, we can reach on the order of 36-49 unit cells containing up to 500,000 atoms.
This should suffice for our purpose of studying random twist angle disorder effects since the disorder is essentially local in nature.
However, to confirm these disorder calculations we consider a related model in Appendix~\ref{appendix:SOC}: a model which can numerically include an order of magnitude more unit cells.
That model has the same features as TBG (the formation of a semimetal miniband and a vanishing velocity at a critical potential strength), confirming the picture presented here.
It is gratifying that we get very similar results in the two models (Appendix~\ref{appendix:SOC}), thus justifying our investigation of twist-angle TBG disorder.

\begin{figure*}
\includegraphics[width=2\columnwidth]{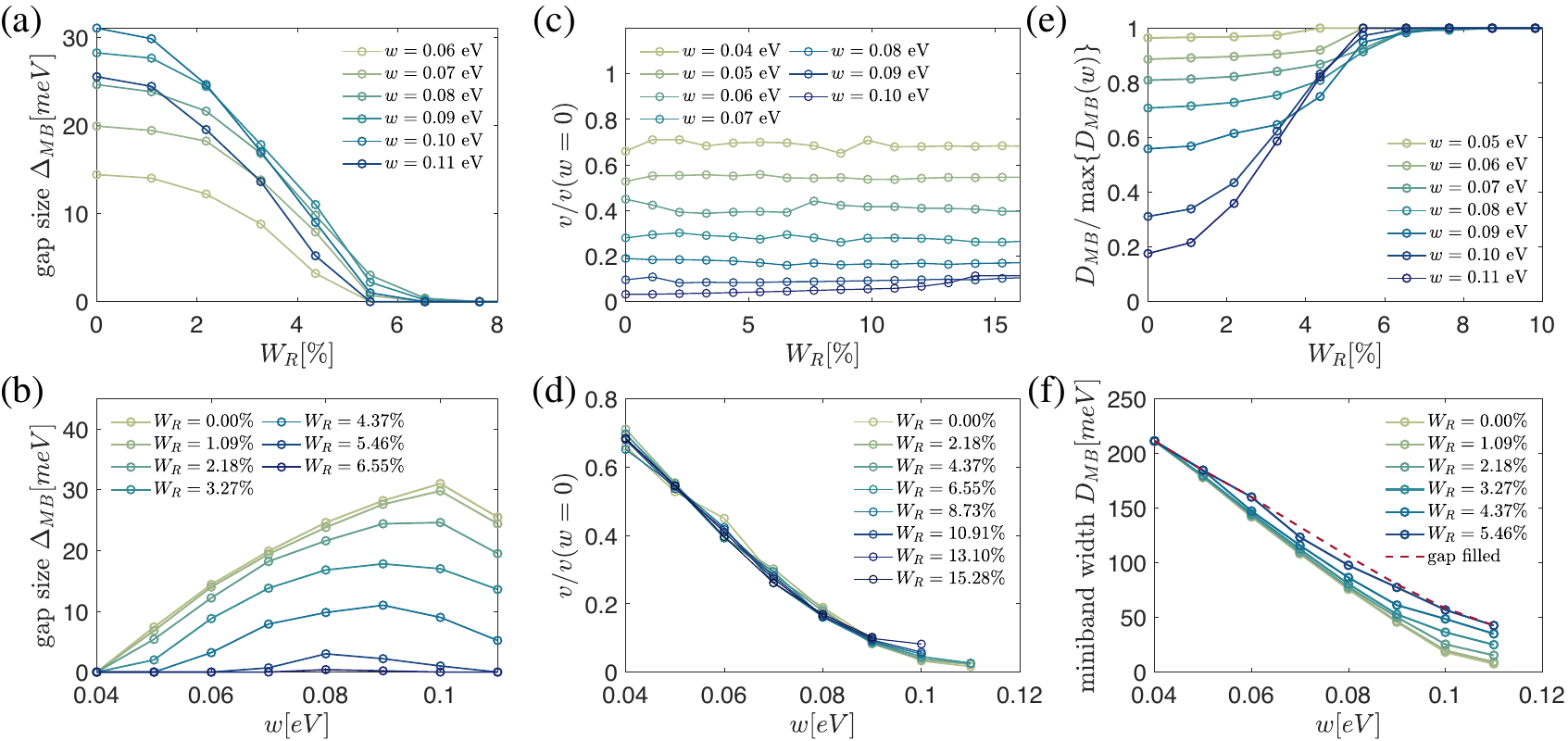}
\caption{(color online) Summary of results on the miniband properties in the TBG model with a clean twist angle $\theta=1.05^{\circ}$ extracted from system sizes $L=569$ and a KPM expansion order $N_C=2^{17}$.
(a,b) The estimated gap size $\Delta_{\mathrm{MB}}$ as a function of disorder strength in the twist angle $W_R$ and the interlayer tunnelings $w$ (where $w=w_1$ and the ratio of AA and BB tunneling to AB and BA tunneling is $w_0/w_1=0.75$).
(c,d) The velocity $v/v(w=0)$ as calculated from the density of states as a function of disorder $W_R$ remains approximately unchanged in the presence of disorder $W_R$ (for each value of $w$).
(e,f) The minibandwidth $D_{\mathrm{MB}}$ for interlayer tunneling $w$ and disorder $W_R$.
Note that for larger disorder strength ($W_R = 6\%$ or above) in (e) the bandwidth appears to plateau; this is just an artifact arising from disorder completely filling out the gap at this point.
While the gap and bandwidth are strongly affected by disorder, the velocity remains unchanged. The red dashed line in (f) that sets the maximum that the minibandwidth can achieve, is determined from the gaps in (b).}
\label{fig:summary_TBG}
\end{figure*}

We focus on the density of states (DOS), that is defined as
\begin{equation}
\rho(E) = \frac{1}{4L^2}\left [\sum_{i}\delta(E- E_i) \right ] \label{eq:DOS}
\end{equation}
where $[ \dots ]$ denotes an average over disorder, phases, and twists in the boundary condition. In what follows we average over $100$ disorder samples.
In order to reach large system sizes we use the kernel polynomial method (KPM) to compute the density of states through an expansion in terms of Chebyshev polynomials and we use the Jackson Kernel to filter out oscillations due to truncating this expansion to an order $N_C$~\cite{Weisse-2006}. In the following, we focus on a linear system of $L=569$ and a KPM expansion order ranging from $N_C=2^{13}$ up to $2^{17}$.
This should give us an essentially exact nonperturbative evaluation of the TBG DOS in the presence of twist disorder.

From the density of states we extract an estimate of the renormalized velocity of the Dirac cones, using the scaling for two-dimensional Dirac cones with velocity $v$,
\begin{equation}
\rho(E) \sim \frac{1}{v^2}|E-E_D| \label{eq:lowEDOS-SM}
\end{equation}
near the Dirac nodal energy $E_D$ and we extract an estimate of $v$ through a fit of the low-energy density of states.
We mention that the Dirac cone approximation is only valid at low TBG energies well below the Van Hove singularities, and hence our extracted Dirac cone velocity applies only at low energies.
Despite the expectation that disorder will induce a small but nonzero density of states at $E_D$, we can still use the scaling in Eq.~\eqref{eq:lowEDOS-SM} to provide an estimate of the renormalized velocity.
To quantify the effect of disorder on the Van Hove peaks in the DOS we make a qualitative estimate of the ``BCS superconducting transition temperature'' from the DOS through
\begin{equation}
T_c \propto \exp\left(-\frac{1}{g\rho(E_{\mathrm{vH}})}\right)
\label{eq:Tc}
\end{equation}
where $E_{\mathrm{vH}}$ is the location of the Van Hove (vH) peak in energy, we take an electron-phonon coupling $g=1$, and $T_c$ is in units of eV for the TBG model.
We stress that we by no means are claiming electron-phonon interaction is the origin of superconductivity in twisted bilayer graphene (although we do not rule out this possibility either). We are only using Eq.~\eqref{eq:Tc} as a qualitative measure of how disorder smears out the Van Hove peaks, which reduces the largest possible mean-field critical temperature in the miniband within BCS theory.
One should think of the effective $T_c$ in Eq.~\eqref{eq:Tc} as a measure of the effective nonperturbative coupling induced by the vH singularity, and Eq.~\eqref{eq:Tc} is a simple quantitative approximation to estimate the effect of twist angle disorder on the vH singularity expressed in units of energy (i.e.\ coupling strength).
The fact that this formula coincides with the BCS formula for the superconducting transition temperature is a matter of convenience in this respect.
Any other such formula should provide the same qualitative results although the quantitative details will depend on the specific form of the chosen formula.

\section{Results}\label{sec:results}

To begin, we first discuss the effects of a random twist angle in the effective TBG lattice model.
Since the twist shows up explicitly in the interlayer tunneling term, randomness appears solely in this part of the Hamiltonian.
However, interlayer tunneling either occurs between equivalent sites or nearest neighbors (on the triangular Bravais lattice) between the two layers. This is due to $\mathcal T_0$ and $\mathcal T_1$ terms in Eq.~\eqref{eq:tbg_ham}, and thus, randomness in the twist angle will induce contributions from both of these terms.

The miniband that is formed due to the twist can be characterized by the following independent and complementary quantities:
(1) the size of the energy gaps (mostly at `higher' energies at the miniband edges) separating it from the rest of the states,
(2) the effective low-energy velocity of the Dirac cones in the minizone,
(3) the minibandwidth, and
(4) the size and shape of the Van Hove peaks (which are strongly enhanced due to the formation of the miniband itself before disorder is taken into account).
These features are all summarized in Fig.~\ref{fig:cleanDOS}.

First, as shown in Fig.~\ref{fig:DOSE_tbg}, disorder destabilizes the integrity of the miniband that is created due to the twist.
When the gaps first develop, they appear at energies $\sim v_\mathrm{F} k_{\mathrm{D}}\sin(\theta/2)$ and their size is perturbatively controlled by $w_1=w$.
As the figure shows, the gaps become soft due to averaging together different patches of random twist angles.
We extract the miniband (MB) gap $\Delta_{\mathrm{MB}}$ for various values of interlayer tunneling ($w_1$) and disorder strength, as shown in Fig.~\ref{fig:summary_TBG}(a,b).
Increasing the interlayer tunneling and approaching the magic-angle condition makes the semimetal miniband more pronounced and stable by increasing the size of the gap, which is maximal near $w=\unit[0.1]{eV}$.
Introducing finite disorder makes these gaps soft and the average band gap fills in monotonically with increasing disorder.
Eventually, the gap is filled in completely, which we find occurs roughly for $W_R = 6 \%$ of the clean twist angle, and there is no longer a clear separation between the miniband and the rest of the states.
This effect is clearly visible in experiments, as the band insulating gap is destroyed (e.g. as seen in Ref.~\cite{Yankowitz-2019}).
The sensitivity of the gap to disorder in the twist angle is rather intuitive, as the location of the gap is dictated by the scattering between the Dirac nodes of equal chirality but different layers, and the energies that mix to open a gap are  determined by $\theta$, whereas the size of this gap is dictated by $w$.
But the fact that the primary insulating gap at the full filling of the moir\'e miniband may be completely suppressed by a twist-angle disorder as small as just $<10\%$ is non-obvious-- naively on perturbative grounds one may expect a relative disorder of the order of unity (i.e.\ 100\%) in order to completely suppress the gap.
Clearly, the miniband insulator is very sensitive to twist-angle disorder, and this may be the reason why the measured gaps vary quite a bit from sample to sample even for nominally fixed twist angles.

Second, we discuss to the features of the miniband which presumably drive strong correlation effects, namely the renormalized Dirac cone velocity $v$ [Fig.~\ref{fig:summary_TBG}(c,d)] and the size of the minibandwidth $D_{\mathrm{MB}}$ [Fig.~\ref{fig:summary_TBG}(e,f)].
Surprisingly, we find that the Dirac velocity is remarkably robust to disorder and while it is strongly suppressed for increasing $w$ (as expected since this is an effective decrease of the twist angle), increasing disorder enough even to fill in the band gaps and suppress $T_c$ completely is not sufficient to modify the effective velocity which maintains its clean value in a robust manner even in the highly disordered situation.
As shown in Figs.~\ref{fig:summary_TBG}(c,d), the effective velocity extracted from Eq.~\eqref{eq:lowEDOS-SM} does not renormalize until the disorder is very large; in particular, Fig.~\ref{fig:DOSE_tbg} demonstrates that the low-energy scaling of the DOS $\rho(E)\sim |E-E_D|$ remains robust for a range of disorder with an unmodified slope.
Close to the magic-angle regime ($w\approx \unit[0.11]{eV}$), the vanishing of the velocity is becoming rounded out; however to see this develop for a large disorder range is challenging as we are limited by the energy resolution needed and therefore we only present results for disorder strengths where the scaling in Eq.~\eqref{eq:lowEDOS-SM} is clearly visible.
In any case, close to $\theta_{\mathrm{Magic}}$, the whole concept of a velocity becomes dubious as the TBG basically is a completely flatband system with essentially no energy regime available for the Dirac cone approximation to apply.

The minibandwidth $D_{\mathrm{MB}}$ is similarly substantially reduced as we approach the magic-angle regime, as shown in Figs.~\ref{fig:summary_TBG}(e,f).
However, disorder both fills in the band gaps [Figs.~\ref{fig:summary_TBG}(a,b)] and also broadens the minibandwidth which we we are able to track provided the band gaps have not completely filled in [that we mark with a red dashed line in Fig.~\ref{fig:summary_TBG}(f)].
The effect of disorder on the minibandwidth is much stronger than the effect on the velocity, and we expect disorder may reduce the strength of correlations by broadening the size of the miniband.
It will not, however, have a very large effect on the Dirac velocity for weak disorder.
We believe that such effects of disorder strongly suppressing correlation effects in the system (by effectively broadening the minibandwidth) are already apparent in the experimental samples since the insulating gaps (i.e. the correlated insulator phase) at commensurate fractional filling of the miniband often do not show up in many samples, and when they do, the typical correlated insulating gap energies are often rather small and vary strongly from sample to sample.

\begin{figure}
\includegraphics[width=\columnwidth]{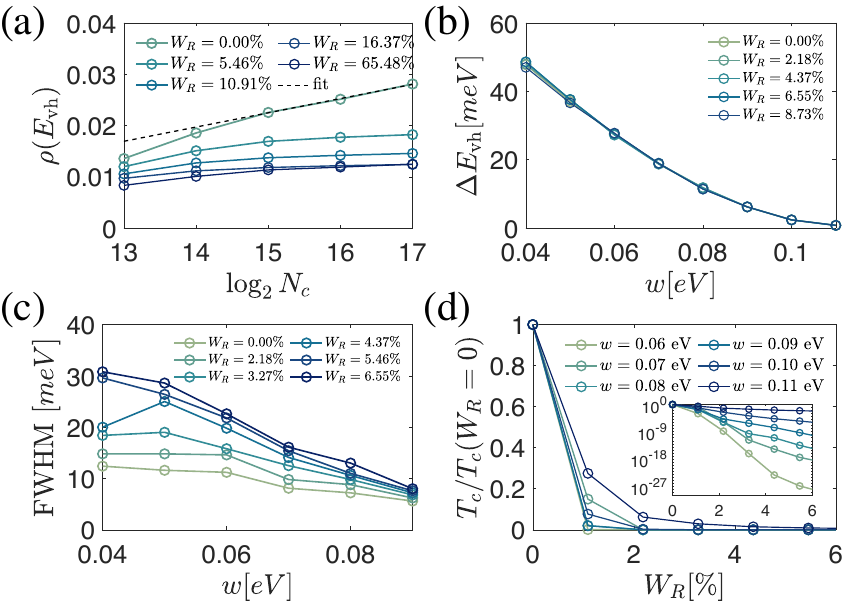}
\caption{The effects of twist disorder on the properties of the Van Hove peaks for a clean twist angle $\theta=1.05^{\circ}$, a linear system size $L=569$, and a varying KPM expansion  order ($N_C$) in (a) whereas in (b,c,d) we use $N_C=2^{17}$.
(a) As we scale the Chebyshev expansion order, we see that the Van Hove peak is logarithmically divergent (with a fit shown as a black dashed line), but once we add disorder, it rounds out and saturates to a finite value.
(b) The energy separation between Van Hove peaks remains stable as disorder increases even though we find (c) that the full-width half-max (FWHM) of the Van Hove peaks becomes broader as disorder increases.
(d)  The estimated BCS critical temperature or the effective coupling constant [see Eq.~\eqref{eq:Tc} in the main text] from the density of states at the Van Hove peak as disorder is tuned up for various values of $w$.
}
\label{fig:Tc_TBG}
\end{figure}

While the gap and hence minibandwidth are strongly affected, disorder also has an effect on the finer features of the minibands.
The effects of twist disorder on the Van Hove peaks are captured quantitatively in Fig.~\ref{fig:Tc_TBG}.
Van Hove singularities in 2D have a logarithmic singularity and thus should diverge with system size $\rho(E_{\mathrm{vH}}) \sim \log L$. However, in our KPM calculations, we expect that the finite expansion order ($N_C$) produces a larger finite size effect than the system size. Therefore, we study the scaling of the Van Hove peaks with the KPM expansion order in Fig.~\ref{fig:Tc_TBG}(a).
This clearly demonstrates that the 2D logarithmic vH singularity, manifesting the scaling $\rho(E_{\mathrm{vH}}) \sim \log N_C$ in the clean limit, becomes rounded out due to  disorder and no longer diverges with increasing $N_C$.
Interestingly, however, the location and separation of the Van Hove peaks is very insensitive to disorder as shown in Fig.~\ref{fig:Tc_TBG}(b).
Despite the average location of the Van Hove peaks remaining fixed, disorder broadens them out as we show in Fig.~\ref{fig:Tc_TBG}(c) by computing the full width at half maximum (FWHM). Not only does this figure demonstrate that the FWHM of the vH peaks strongly decreases with increasing $w$ it also shows that the effects of disorder on the vH peaks are much stronger for smaller $w$ away from the magic-angle regime.
This subtle effect of twist angle disorder on the vH peaks is rather non-obvious.

To study disorder effects on the Van Hove peaks in more detail we extract an estimate of the mean-field BCS superconducting transition temperature from Eq.~\eqref{eq:Tc} due to the DOS at the Van Hove peak energy.
We show the effects of interlayer tunneling and disorder on $T_c$ in Fig.~\ref{fig:Tc_TBG}(d). Since the Van Hove peaks are strongly affected by $w$, we normalize $T_c$ by its value in the clean limit to compare our disordered results for each value of $w$.
In the absence of randomness, shown in Fig.~\ref{fig:cleanDOS_TBG}(a), as we increase $w$ the minibandwidth shrinks, pushing the same number of states down to a lower energy scale, which in turn enhances the Van Hove peaks considerably.
Upon introducing the twist disorder, $T_c$ is suppressed, and this effect, rather unexpectedly, is most pronounced for weak interlayer tunneling strengths, whereas for $w$ close to the magic-angle condition ($w = \unit[0.11]{eV}$ for $\theta \approx 1.05^{\circ}$) we find $T_c$ is not as strongly affected by weak disorder in comparison.


\section{Discussion}
\label{sec:discussion}

First, we discuss our approximations in incorporating twist-angle disorder effects in the otherwise defect and impurity free clean twisted bilayer graphene.
Using an effective model for twisted bilayer graphene we have theoretically investigated effects of twist angle disorder nonperturbatively by breaking the system into four separate equally sized squares each with a random twist angle around a mean value of $\theta_0 = 1.05^{\circ}$ close to the magic angle.
To understand the effects of our choice of modeling twist disorder with four equal sized squares, in Appendix~\ref{appendix:SOC} we analyze a simpler model to determine the effects of this patching scheme.
By breaking the system into  randomly sized rectangles with each having a different twist value we show that our qualitative results are  robust.
Increasing the patch number as well as changing the size and shape introduces more randomness into the system and increases the effective disorder strength overall.
Therefore, the  amount of randomness in each sample  is a function of both the random distribution and the number of patches.
Here, we want to ensure that each patch has enough sites in it to host a well defined low-energy semimetallic miniband at the magic-angle regime ($w=0.11$ eV and $\theta\approx 1.05^{\circ}$) and therefore have focused on 4 squares and total linear system size $L=569$ (in terms of Bravais lattice sites).
Increasing the number of squares or modifying the shape will only introduce more randomness into the system.

We have introduced an effective lattice model of twisted bilayer graphene that is local, only requiring nearest neighbor (on the triangular Bravais lattice) interlayer hopping terms which already captures many of the features of the continuum model~\cite{lopesdossantos_graphene_2007,BistritzerMacDonald2011}, such as the miniband gaps, Van Hove peaks, as well as the velocity and minibandwidth renormalization.
The model we have used  maintains the $C_2 T$ symmetry but breaks the $C_3$ symmetry.
As a result, the velocity and minibandwidth renormalization are affected in the magic-angle regime, which leads to a non-vanishing velocity (see Appendix~\ref{appendix:pt}) and an overestimate of the minibandwidth.
Moreover, the model also introduces fine structure into the Van Hove peaks that we attribute to additional zone folding that appears in the lattice model.
Despite these shortcomings, this lattice model does capture the qualitative behavior of the low energy miniband very well while remaining local and easy to work with numerically.
It is possible to construct an effective lattice model that preserves the $C_3$ symmetry and more accurately reproduces the continuum model in the magic-angle regime with a true vanishing velocity.
However, this requires  a more non-local interlayer hopping model keeping  up to third nearest neighbor tunneling terms on the triangular lattice, which will appear in Ref.~\cite{Fu-2018} (our conclusions change little using this more sophisticated model).
In experiments on twisted bilayer graphene, the encapsulating substrate as well as other forms of disorder break the $C_3$ symmetry explicitly.
Therefore, we do not expect that the weak breaking of this symmetry in our effective lattice model affects our conclusions on the qualitative experimental implications of disorder in the twist angle.

Now we briefly summarize our main findings.
Our results clearly demonstrate that the low-energy scaling of the semimetal miniband $\rho(E) \sim v^{-2}|E-E_D|$ and the effective Dirac cone velocity ($v$) are remarkably robust to disorder in the twist angle.
While this robustness slightly weakens in the magic-angle regime due to disorder eventually rounding out the velocity minimum, we find that  $v$ is essentially disorder independent for less then 15\% of randomness in the twist angle.
This result suggests that the semimetallic scaling near the magic-angle regime should be clearly visible in transport experiments
that average over the whole sample.
Indeed, our findings are consistent with the experimental observations on twisted bilayer graphene that have observed a robust ``V-shaped'' conductance minimum at charge neutrality~\cite{Yao2018,Yankowitz-2019} that signifies that the semimetallic low-energy scaling persists in spite of the inevitable presence of twist angle fluctuations in the sample.
The existence of a low-energy Dirac cone is protected against twist angle disorder.
It is interesting to note that twisted bilayer graphene samples that are ``massaged'' to remove bubbles that have formed in the ``tear and stack'' approach exhibit an insulating phase at charge neutrality~\cite{Lu-2019}.
Presumably, this procedure eliminates some of the twist disorder in the sample and, as a result, domains of twist angle that still possess the semimetallic density of states no longer contribute to the density of states near $E=0$.
Thus, the suppression of twist disorder comes with the price of a strong modification of the observed density of states at low energies.

On the other hand, the minibandwidth is much more strongly affected by disorder, and $D_{\mathrm{MB}}$ monotonically increases for increasing disorder strength until the gap is completely filled in and the integrity of the miniband is lost.
Similarly, we have found that the insulating gap that separates the miniband from the rest of the states is completely filled in at weak disorder strengths ($\sim 6 \%$ of the clean twist angle).
This strong sensitivity of the single-particle gaps to twist disorder has been observed in Ref.~\cite{Yankowitz-2019} by placing leads at different places in the sample and finding very strong variations in the gap energies.
We suspect that twist disorder will have an even stronger effect on the gaps at the correlated insulator filling fractions.
In particular, the increase of the effective minibandwidth by twist disorder entails an effective lowering of the dimensionless correlation strength (i.e. the effective $U/t$ value in the Hubbard-type models) since the Coulomb interaction energy (i.e. the effective $U$) should not be affected by the disorder whereas the minibandwidth (i.e. the typical $t$) increases.
These combined results imply that disorder will reduce the strength of many-body correlations by increasing the bandwidth of the miniband but will not affect the flatness of the Dirac cones.
This interesting subtle prediction of our nonperturbative theory may already have support in the existing experiments since many otherwise high-quality TBG samples (i.e. made from extreme high-mobility graphene sheets) often manifest correlated insulating phases that are very weak, and it is unclear why the correlated insulator phase at commensurate fractional fillings is not universally seen in all TBG samples of nominally same quality at the same twist angle.
We propose that the twist angle randomness is responsible for causing sample to sample variations in the TBG physics for the same average twist angles.

Last, the Van Hove peaks  are a clear signature of the miniband in twisted bilayer graphene  experiments~\cite{li_observation_2010,Luican-2011,brihuega_unraveling_2012,wong_local_2015,kerelsky_magic_2018,choi_imaging_2019,jiang_charge-order_2019,xie_spectroscopic_2019}. Our results demonstrate that the location of the Van Hove peaks of the miniband as well as their separation in energy, which is minimized in the magic-angle regime, are essentially unaffected by twist-angle disorder.
Twist-angle randomness smears out the logarithmic Van Hove singularity without affecting their locations in energy.
As a result, the density of states becomes an analytic function of energy and system size at the Van Hove peaks in the presence of twist-angle disorder.
We have qualitatively assessed the impact of disorder on  the mean-field BCS superconducting transition temperature in the miniband by considering  a Fermi energy at a Van Hove peak. We have found that twist disorder strongly suppresses $T_c$ [as it is defined in Eq.~\eqref{eq:Tc}]. If the superconductivity in twisted bilayer graphene is BCS like then our results suggest that samples with large amounts of disorder in the twist angle will likely not superconduct.
This is again consistent with experimental observations where not all samples with similar twist angles manifest superconductivity, and we speculate that this nonuniversality is connected with the presence of variable twist-angle disorder in different samples.

\section{Conclusion}
\label{sec:conclusion}

In this work we construct an effective lattice model of twisted bilayer graphene which we use to study the effects of disorder in the twist angle within a nonperturbative essentially exact theory.
We investigate how our choice of modeling disorder affects our results through a detailed investigation of a related but simpler model in Appendix~\ref{appendix:SOC}. 
It will be interesting in future work to incorporate larger and smoother domain walls between different twist angles than we have considered here.
We demonstrate how randomness in the twist angle affects various properties of the low energy miniband through numerically exact calculations of the density of states using the kernel polynomial method.
Remarkably, we show that the velocity of the Dirac cone is robust to disorder, whereas the other features of the miniband are rather sensitive to randomness in the twist angle.
Last, we also discuss how the implications of our theory might already been observed out in existing experimental data and have given guidance for how these disorder effects can be used to help understand the putative strongly correlated effects seen in experiments.

\acknowledgements{We thank Shaffique Adam, Eva Andrei, Zhen Bi, Jason Ho, Philip Kim, Elio K\"onig, and Alex Thomson  for useful discussions. JHW, SDS, and JHP acknowledge the Aspen Center for Physics, where some of this work was completed, which is supported by National Science Foundation Grant No. PHY-1607611. JHP and SDS also acknowledges the Kavli Institute for Theoretical Physics, which is supported by NSF Grant No. PHY-1748958, where the early stages of this work were performed.
SDS is supported by the Laboratory for Physical Sciences.
 The authors acknowledge The University of Maryland supercomputing resources (http://hpcc.umd.edu), the Beowulf cluster at the Department of Physics and Astronomy of Rutgers University, and the Office of Advanced Research Computing (OARC) at Rutgers, The State University of New Jersey (http://oarc.rutgers.edu) for providing access to the Amarel cluster and associated research computing resources that have contributed to the results reported here.

}

\appendix

\section{Spin-orbit coupling model}
\label{appendix:SOC}
\begin{figure}
\includegraphics[width=\columnwidth]{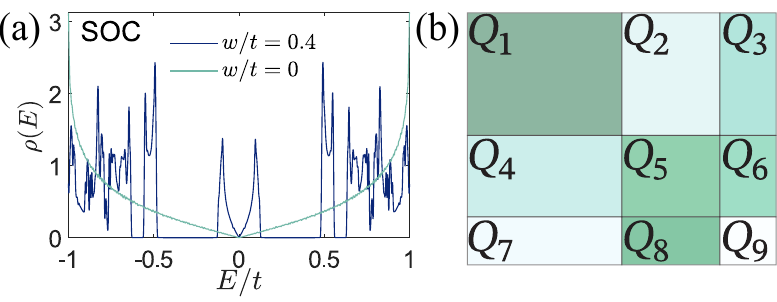}
\caption{(color online) (a) The calculated density of states $\rho(E)$ as a function of energy $E$ for the spin-orbit coupled model of Dirac points perturbed by a quasiperiodic potential, with a quasiperiodic wavevector $Q=2\pi F_{n-2}/F_n$ with the system size $L=F_n=144$ and a KPM expansion order $N_C=2^{14}$.  (b) A depiction of how we break up the SOC square lattice model into regions of different quasiperiodic wavevector $Q_i$ (to simulate disorder), which are taken from a box distribution about a central value. We vary both the number of regions and the size of disorder in each region.}
\label{fig:cleanDOSSOC}
\end{figure}
In addition to the lattice model of twisted bilayer graphene (described in the main text), the second disordered TBG-like model we study is  a two-dimensional tight-binding model with
spin-orbit coupling (SOC) in the presence of a quasiperiodic potential, which is defined as \cite{Fu-2018}
\begin{equation}
H_{SOC} =  \frac{1}{2}\sum_{\br,\mu = x,y}  (it \chi_{\br}^\dagger \sigma_\mu \chi_{\br + \hat \mu} + \mathrm{h.c.})+ \sum_{\br}   V(\br) \chi_{\br}^\dagger \chi_{\br} .
\end{equation}
where $t$ is the hopping strength, the lattice spacing is set to unity, $\chi_{\br}$ denotes a two component spinor of annihilation operators, and $\sigma_{\mu}$ are the Pauli operators. We mimic the effect of a twist through a quasiperiodic potential
\begin{equation}
V(\br) =W \sum_{\mu = x,y}  \cos(Q r_{\mu}+ \phi_\mu) ,
\end{equation}
of strength $W$, an incommensurate (or quasiperiodic) wave-vector $Q$, and $\phi_{\mu}$ is a random phase sampled between $0$ and $2\pi$.
We average over twisted boundary conditions to reduce the finite size effects.
The goal of using this second model is to test the universality of the conclusions we reached in the main text using the TBG lattice model. Note that for the DOS computed in the SOC model we normalize the DOS in Eq.~\eqref{eq:DOS} by a factor of $2L^2$ as opposed to $4 L^2$ to account for the smaller local Hilbert space.

\begin{figure}[h]
\includegraphics[width=0.95\columnwidth]{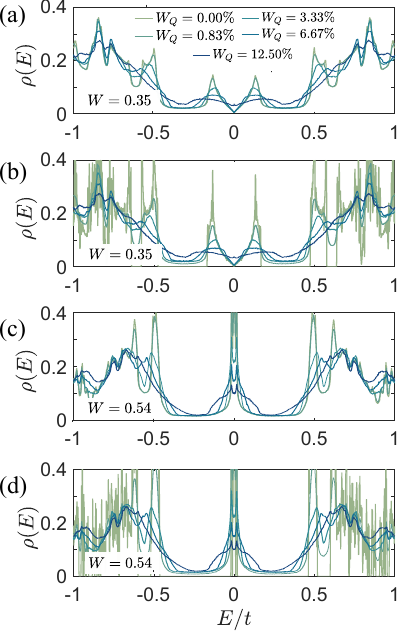}
\caption{The disorder-free density of states $\rho(E)$ as a function of energy $E$ obtained from a linear system size $L=144$ and a KPM expansion order $N_C=2^{14}$ starting in the semimetal regime of the model, comparing the case of a fixed random phase across the entire sample (b, d) and a different random phase in each patch (a, c) for different strengths of disorder in the wavevector and $n_P=7$ randomly placed patches. Note that the random phase in each patch is disordered even for $W_Q=0$. }
\label{fig:DOS_rp}
\end{figure}

The effect of the quasiperiodic potential on Dirac points is similar to twisting two layers of graphene~\cite{Fu-2018}. As shown in Fig.~\ref{fig:cleanDOSSOC}(a) for a large enough value of $W$, a semimetal miniband forms with a renormalized velocity, sharp Van Hove peaks, and hard gaps separating it from the rest of the spectrum. Importantly, the SOC model has the great advantage that the formation of minibands, a hard gap, and flat bands are clearly visible on much smaller system sizes compared with the effective lattice model of TBG. Here, it is sufficient to consider system sizes of $L=144$ or larger to see these effects, whereas in the TBG model the minimum number of sites required to form a clear miniband is at least a linear system size of $L=300$. 

In the calculations of the TBG model we broke the system into four squares of equal size, which was for simplicity of modeling while being able to correctly capture the formation of the miniband by keeping each patch sufficiently large.
We now investigate the effects of making the size and shape of these regions random as well as increasing the number of random patches $n_P$, something that the computational demands of the lattice TBG model did not allow us to do.
We divide the $L\times L$ lattice into $(n_P)^2$ domains, by cutting it through $n_P-1$ vertical and $n_P-1$ horizontal lines which are randomly located.
Each domain $i$ is given a quasiperiodic wavevector and phase $[Q(i), \phi_\mu(i)]$, as illustrated in Fig.~\ref{fig:cleanDOSSOC} (b).
We introduce randomness in $Q$ in a similar way as in the main text, such that $Q(i)=Q_0 +\delta Q_i$ where $Q_0=2\pi F_{n-2}/F_n$, $F_{n}$ is the $n$th Fibbonnaci number, and we take the system size $L=F_n$ so that $Q_0$   is a rational approximant to the irrational number $2\pi(2/[\sqrt{5}+1])^2$. In each domain (or patch) $\delta Q_i$ is taken from a uniform distribution around $Q_0$, i.e. $Q(i) \in [(1-W_Q)Q_0,(1+W_Q)Q_0]$ and $W_Q$ is expressed as a percent (similar to the random disorder $W_R$ in the main text).
For the results on the SOC model we average over 300 disorder samples.

\begin{figure}
\includegraphics[width=0.95\columnwidth]{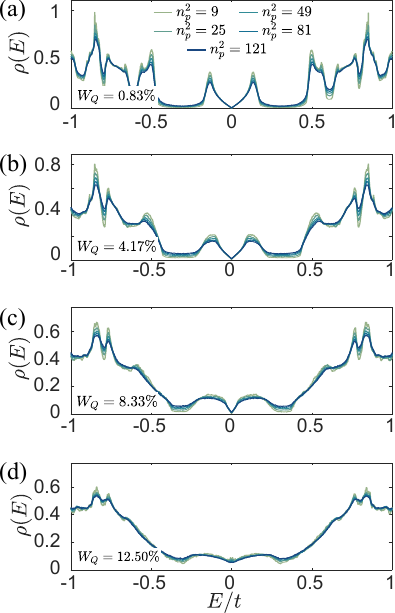}
\caption{(color online) Density of states as a function of energy in the semimetallic regime of the SOC model focusing on the miniband at low energy using a linear system size $L=144$ and a KPM expansion order $N_C=2^{14}$. We focus on the effects of the different number of random patches used for various different disorder strengths in the quasiperiodic wavevector $W_Q$ from $W=0.35$. Here we are taking one global phase across the sample to isolate the effects of randomness in $Q$ and choice of patches alone.  }
\label{fig:fixedphiSM}
\end{figure}

\begin{figure}
\includegraphics[width=0.95\columnwidth]{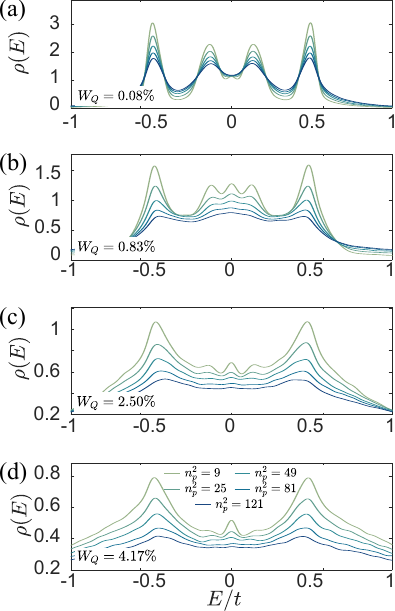}
\caption{(color online) Density of states as a function of energy in the magic-angle regime ($W=0.54$) of the SOC model focusing on the miniband at low energy with a linear system size $L=144$ and a KPM expansion order $N_C=2^{14}$. We are displaying the effects of  different number of patches of a random wave vector across the sample. }
\label{fig:fixedphiM}
\end{figure}

\begin{figure}
\includegraphics[width=\columnwidth]{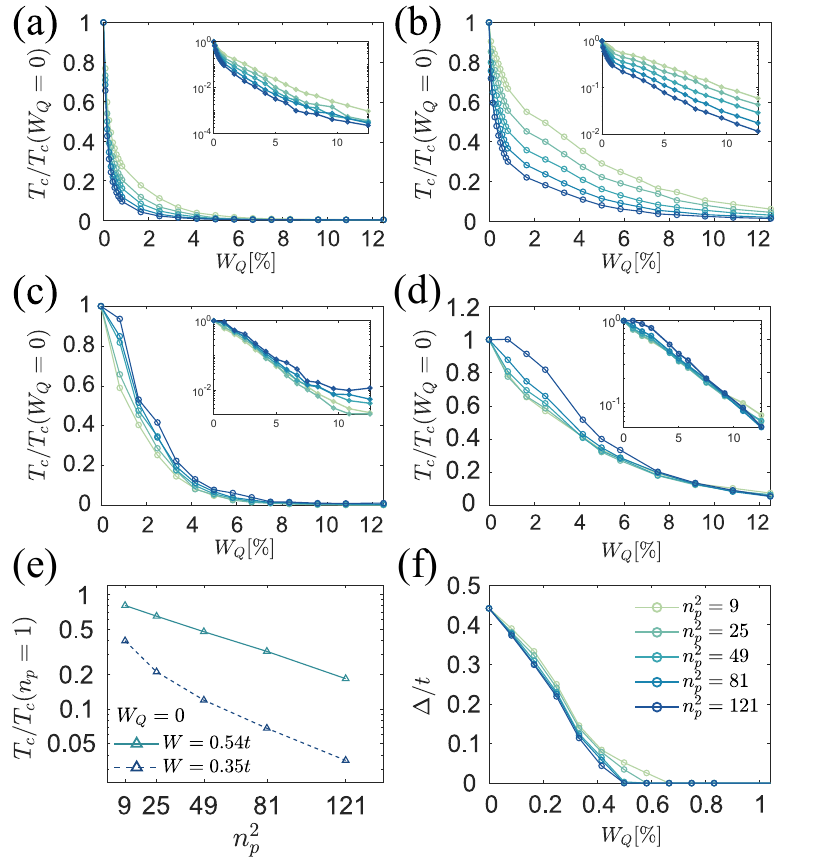}
\caption{(color online) The estimated critical temperature (or effective coupling-- see Eq.~\eqref{eq:Tc} in main text) from the Van Hove peaks in the DOS as a function of randomness in the twist vector comparing two choices for the random phase for different number of randomly placed patches.
(Top row) One fixed phase, corresponding to a single rotation origin. (Middle row) Random phases $\phi_{\mu}(i)$ in each block. The left panels are $W=0.35$ in semimetal phase, while the right panels are $W=0.54$ at the magic-angle.
Random phases in each block produce very strong randomness in the model and smears out the Van Hove peaks more easily.
(Bottom left) The critical temperature $T_c$ with random phases $\phi_{\mu}(i)$ in each block but without randomness in $Q$, as function of number of patches $n_p^2$, and normalized by $T_c$ with only one patch.   (Bottom right) The gap size as function of randomness. Comparing to the suppression of $T_c$, the gap is filled in for $W_Q \approx 0.5\%$, which is much smaller than the critical $W_Q$ ($\sim 10\%$) needed for Van Hove peaks to be smeared out. These results are obtained from data using a linear system size $L=144$ and a KPM expansion order $N_C=2^{14}$.}
\label{fig:TCSOC}
\end{figure}

In order to understand the role of taking a uniform phase [$\phi_j$ in Eq.~\eqref{eq:Hopping-Ts}] in the TBG calculations we consider choosing the phase  in each patch $\phi_{\mu}(i)$ in two distinct ways, which are:
(A)~One global phase $\phi_\mu(i)=\phi$, which is equivalent to our set up in the TBG model.
(B)~In each patch, the phases $\phi_{\mu}(i)$ are independently picked from a uniform distribution $[-\pi, \pi]$, which amounts to a disorder potential even for a fixed wavevector across the sample.
Option (A) has no discontinuity in the phase across  the boundaries of each patch. Note that because of the
variation in $Q$, even fixing $\phi_\mu(i)=\phi$ to be one global phase does not enforce a
continuous boundary condition across patches.
The most random choice we can make is through option (B), which means any phase can be chosen on each patch, with no restriction.
In this case there is a sharp jump of the potential across all of the patches.
In particular, when the number of domains approaches the number of sites, the quasiperiodic potential turns into random disorder potential.
The ``randomness'' of option (B) is clearly the strongest  and is not controlled by the parameter $W_Q$. This is demonstrated in Fig.~\ref{fig:DOS_rp}, which shows that randomness in the phase smears out the fine features of the density of states and fills the gaps in more easily and is qualitatively similar to the case with  a fixed phase.
Thus, randomness in the phase is not essential to include to study disorder, and in the following we will mainly focus on keeping the phase fixed throughout the sample.

To understand the effects of a finite number of patches we present results in the semimetal ($W\approx 0.35t$) and magic-angle ($W\approx 0.54t$) regime of the SOC model (see Ref.~\cite{Fu-2018}) in Figs.~\ref{fig:fixedphiSM} and \ref{fig:fixedphiM}. A clear trend in all of the results is that  increasing the patch number enhances the randomness, which effectively increases the strength of disorder. This is demonstrated in Fig.~\ref{fig:TCSOC} by the gaps becoming soft for weaker disorder strength, as well as an increased rounding of the Van Hove peaks  as we increase the number of random patches. Eventually, at large enough disorder in the wavevector, any remnant of the semimetal scaling regime is destroyed, as shown in Fig.~\ref{fig:fixedphiSM}. In the magic-angle regime as shown in Fig.~\ref{fig:fixedphiM}, which has a small miniband and a large density of states at the Dirac node energy, we find that disorder systematically broadens the size of the minibandwidth while also smearing out the structure of the DOS at finite energy. Similarly, increasing the number of random patches effectively increases the strength of disorder.

We capture the effects of disorder on the Van Hove peaks through $T_c$ [see Eq.~\eqref{eq:Tc} in the main text for the definition of $T_c$, which is simply an effective coupling constant inspired by the BCS theory], which is shown in Fig.~\ref{fig:TCSOC} for wavevector disorder. We find that disorder reduces $T_c$ monotonically, however when compared to the main insulating gap isolating the miniband we find that the Van Hove peaks are relatively much more robust than the main miniband gap. This features is distinct from what we saw in the case of TBG in the lattice model (main text), where $T_c$ was suppressed more strongly than the gap.
Given this dichotomy, we believe that the lattice model should be trusted more in capturing the Van Hove physics of real TBG, and thus, $T_c$ is likely to be suppressed more than the main insulating gap in the presence of TBG twist disorder.

\begin{figure}
\includegraphics[width=\columnwidth]{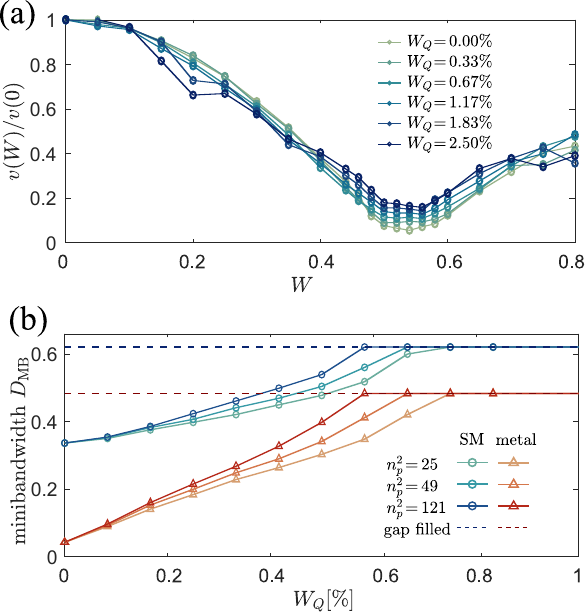}
\caption{Effects of disorder on the renormalization of the velocity of the Dirac cone and the minibandwidth using a linear system size $L=144$ and a KPM expansion order $N_C=2^{14}$. (a) Effective velocity of the Dirac cone and how it is rounded out due to randomness in the wavevector. The finite velocity in the magic-angle regime for $W_Q=0$ is just a finite size effect~\cite{Fu-2018}. (b) Minibandwidth as a function of disorder in the quasiperiodic wavevector, which monotonically broadens for increasing disorder until the gap is filled in and the miniband is no longer separated from the rest of the band (marked as dashed lines). We include both $W=0.35$ for semimetallic phase and $W=0.54$ for the magic-angle regime. Note that we have set $t=1$ here.}
\label{fig:SOC_velocity}
\end{figure}

We now turn to the effects of wavevector disorder on properties of the Dirac velocity and the minibandwidth, as shown in Fig.~\ref{fig:SOC_velocity}. The velocity that vanishes in the magic angle regime is rounded out and remains finite due to the finite disorder strength. Away from the magic angle regime, the effects of disorder on the velocity remain weak. Moreover, the minibandwidth broadens with both increasing disorder strength and the number of patches until the gaps are completely filled. This is consistent with the behavior of the TBG model in the main text, namely that twist disorder weakly effects the velocity and increases the size of the minibandwidth, and the latter effects weakens the strength of correlations in the miniband.
Thus, both models predict a universally robust disorder-resistant Dirac cone velocity at low energies and a considerable disorder-induced broadening of the minibandwidth, thus weakening the correlated insulator phase.

\section{Perturbation theory for the lattice model}
\label{appendix:pt}

It is useful to understand the result of second-order perturbation theory in interlayer tunneling strength within our lattice model to see how the vanishing of velocity is modified near the magic-angle due to the $C_3$ symmetry breaking.
This is done to analytically establish the accuracy of our lattice model of the main text compared with the TBG continuum model of Ref.~\cite{BistritzerMacDonald2011}.

This begins with diagonalizing the free part of the Hamiltonian Eq.~\eqref{eq:freeH}.
In first-quantized notation in momentum $({\bf k})$ space this reads simply as
\begin{equation}
    H_0(\mathbf k) = t[(1+e^{i \mathbf k\cdot \mathbf a_1} +e^{i\mathbf k \cdot \mathbf a_2})\sigma^+ + \mathrm{h.c.}].
\end{equation}
The full Hamiltonian is given by $H=H_0 + \mathcal{T}$ and if we translate Eq.~\eqref{eq:Hopping-Ts} into $k$-space matrix elements, we get
\begin{multline}
    \mathcal T = \sum_{j=1}^3 e^{-i\phi_j} T_j f(\mathbf k) \ket{\mathbf k - \mathbf q_j/2}\bra{\mathbf k + \mathbf q_j/2} \\ + \sum_{j=1}^3 e^{i\phi_j} T_j^* f(-\mathbf k) \ket{\mathbf k + \mathbf q_j/2}\bra{\mathbf k - \mathbf q_j/2},
\end{multline}
where $f$ is the real-valued function
\begin{equation}
    f(\mathbf k) = \frac12 + \frac{i}{6\sqrt{3}} \sum_{j=1}^6 (-1)^{j-1} e^{i \mathbf k \cdot \mathbf a_j}.
\end{equation}

We then want to isolate the $K$ point in order to perform perturbation theory around the Dirac cone.
We first note that the bare velocity is $v_0 = \tfrac32 t$ for this cone, and we do the perturbation theory by using Dyson's equation for the Green function
\begin{equation}
    G(\omega, \mathbf k) = \frac1{\omega - H_0(\mathbf k) - \Sigma(\omega,\mathbf k)}.
\end{equation}
At second order in perturbation theory, the self energy $\Sigma(\omega,{\bf k})$ is given by
\begin{multline}
    \Sigma(\omega,\mathbf k) = \underbrace{\sum_{j=1}^3 T_j^\dagger G_0(\omega,\mathbf k - \mathbf q_j) T_j f(\mathbf k - \mathbf q_j/2)^2}_{\text{support near $K$}} \\
    + \underbrace{ \sum_{j=1}^3  T_j^T G_0(\omega,\mathbf k + \mathbf q_j) T_j^* f(-\mathbf k - \mathbf q_j/2)^2}_{\text{support near $K'$}}.
\end{multline}
It is important to notice that while there are six terms here in contrast to the continuum model which has only three.
Those corresponding three have support near $K$ while the rest will give small or negligible curvature corrections (which we will nonetheless account for).

In order to do the perturbative expansion, we identify the small parameters controlling the expansion for the continuum model, these are
\begin{equation}
    \alpha_j = \frac{w_j}{v_{\mathrm F} k_\theta}, \quad j=0,1,
\end{equation}
where $v_{\mathrm{F}} = \frac32 t$.
In Ref.~\cite{BistritzerMacDonald2011} $\alpha_1=\alpha_0=\alpha$ while we will keep them arbitrary to account for lattice relaxation.
To identify curvature corrections, we can further expand in $k_\theta$, so we will have terms that go as $\alpha_j$, $\alpha_j k_\theta$, and $\alpha_j k_\theta^2$.

\begin{figure}
    \centering
    \includegraphics[width=\columnwidth]{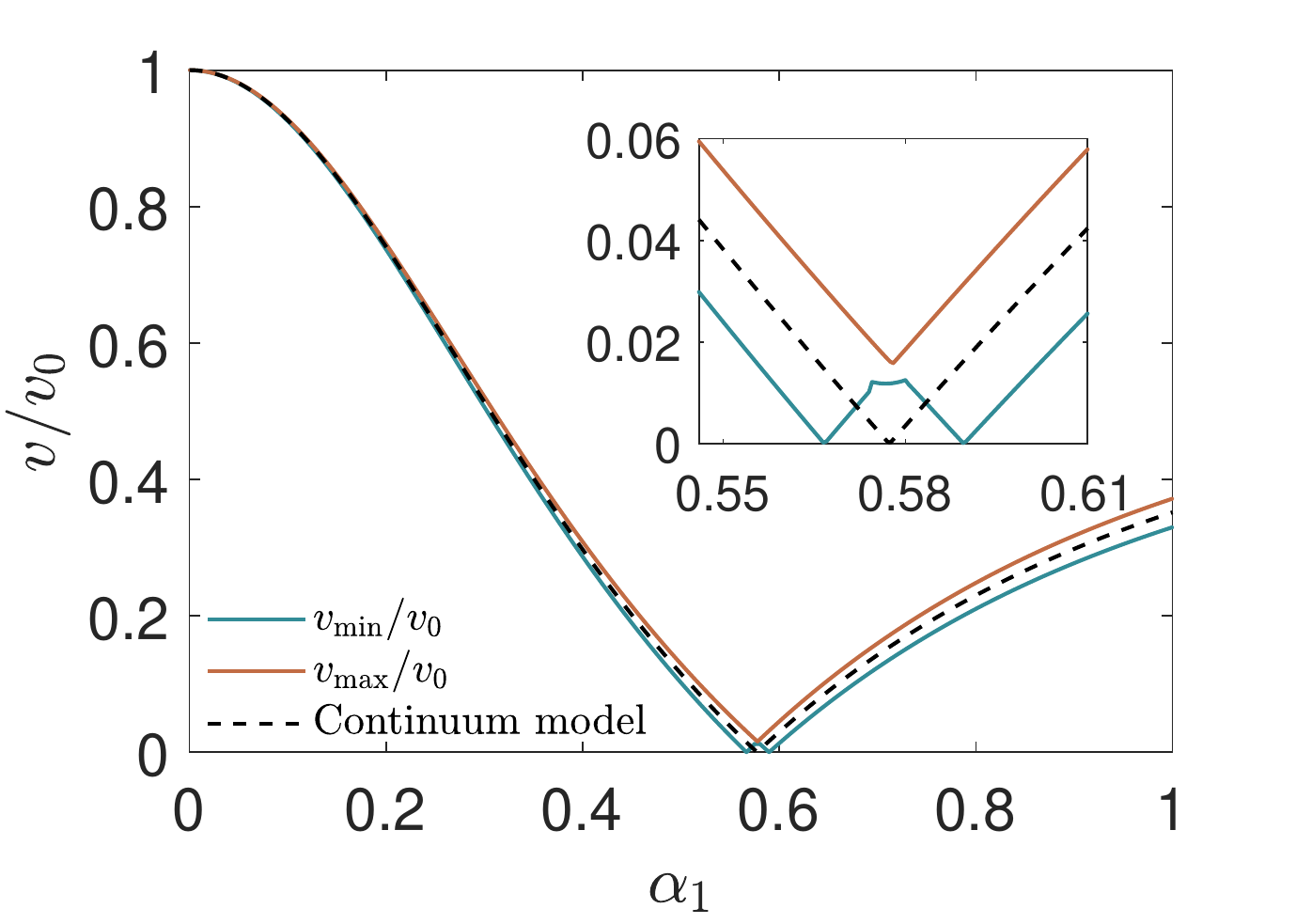}
    \caption{The calculated velocity renormalization in second-order perturbation theory.
    The dashed line is the result from Ref.~\cite{BistritzerMacDonald2011} while the upper and lower curves are the maximum and minimum velocities of the lattice model written down in Eq.~\eqref{eq:tbg_ham} near the $K$ and $K'$ points.
    Notice that the velocity for all states never vanishes as it does in the continuum model even though we have demonstrated in the main text that the density of states matches very well.
    This is plotted for the situation considered in the main text where $\alpha_0 = 0.75 \alpha_1$ and $\theta = 1.05^\circ$.}
    \label{fig:velocity_renormalization}
\end{figure}

Expanding $\Sigma(\omega, \mathbf k)$ for small $\omega$ and $\mathbf k$ and for small curvature, we obtain the following
\begin{widetext}
\begin{multline}
    \Sigma(\omega,\mathbf k)/t \approx \underbrace{-3(\alpha_0^2 + \alpha_1^2)\omega/t}_{\text{WFcn Renorm.}} + \underbrace{\tfrac94 \alpha_0 \alpha_1 k_\theta^2 -\tfrac94(\alpha_0^2 + \alpha_1^2) k_\theta^2 \sigma_x}_{\text{Shift cone}} + \underbrace{\tfrac94 \alpha_0 \alpha_1 \left(- \tfrac32 k_\theta^2 k_x - 2 k_\theta k_y\right)}_{\text{Tilt cone}} \\
    \underbrace{- \left( \tfrac92 \alpha_1^2 - \tfrac{27}{16}\alpha_0^2 k_\theta^2\right) k_x \sigma_x + \tfrac{9}4\alpha_1^2 k_\theta k_x\sigma_y^*   -  \left( \tfrac92 \alpha_1^2 - \tfrac{81}{16}\alpha_0^2 k_\theta^2\right) k_y \sigma_y^* + \tfrac94(2\alpha_0^2 - \alpha_1^2) k_\theta k_y \sigma_x}_{\text{Velocity renormalization}}.
\end{multline}
\end{widetext}
The first term is labeled ``WFcn Renorm.'' for ``wavefunction renormalization.''
The next term labeled ``Shift cone'' is second order in curvature and it shifts both the position of the cone in $k$-space (recall that $C_3$ is broken, so this term is expected) and in energy.
At first-order in the curvature, we obtain the next term labeled ``Tilt cone,'' that acts as a Galilean boost to the cone, tilting it over in $k$-space.
And finally, we obtain corrections labeled ``Velocity renormalization'' since it directly modifies the $v_0 \mathbf k\cdot \bm \sigma^*$ term in the Hamiltonian near the $K$ point.
To obtain the effective Hamiltonian, we put the Green's function in the form
\begin{equation}
  G(\omega, \mathbf k) = \frac{Z}{\omega  - H_{\mathrm{eff}}(\mathbf k)},
\end{equation}
where $Z=[1 + 3(\alpha_0^2 + \alpha_1^2)]^{-1}$ is the quasiparticle residue, and from this we find near the $K$ point
\begin{equation}
    H_\mathrm{eff} = \bm \sigma^* \cdot (V \mathbf k) + \mathbf h_0 \cdot \mathbf k +\frac94 k_\theta^2 t \frac{ \alpha_0 \alpha_1 - (\alpha_0^2 + \alpha_1^2)  \sigma_x}{1+3(\alpha_0^2 + \alpha_1^2)},
\end{equation}
where we have defined
\begin{equation}
    \begin{aligned}
        \mathbf h_0 & = -\frac{9}{8} t \frac{\alpha_0\alpha_1 k_\theta}{1+3(\alpha_0^2 + \alpha_1^2)} ( 3 k_\theta, 4), \\
        V  & =v_0 Z \begin{pmatrix} 1 -  \left( 3 \alpha_1^2 - \tfrac{9}{8}\alpha_0^2 k_\theta^2\right) & \tfrac32(2\alpha_0^2 - \alpha_1^2)k_\theta \\ \tfrac32 \alpha_1^2 k_\theta & 1 -   \left( 3 \alpha_1^2 - \tfrac{27}{8}\alpha_0^2 k_\theta^2\right)  \end{pmatrix}.
    \end{aligned}
\end{equation}
To find the renormalized velocity, consider the velocity operator
\begin{equation}
    \hat{\mathbf v} = \nabla_{\mathbf k} H_{\mathrm{eff}} = V^T \bm \sigma^* + \mathbf h_0,
\end{equation}
and if we take the expectation value with respect to eigenvalues of $H_{\mathrm{eff}}$, we obtain $\braket{\bm \sigma^*} = \mathbf w$ for a normalized vector $\mathbf w = (\cos\vartheta, \sin\vartheta)$.
This allows us to define a velocity
\begin{equation}
    v(\vartheta) = |\braket{\hat{\mathbf v}}| = |V^T\mathbf w(\vartheta) + \mathbf h_0|.
\end{equation}
We can define from this a maximum and minimum velocity
\begin{equation}
    \begin{split}
        v_{\mathrm{min}} & = \min_\vartheta v(\vartheta), \\
        v_{\mathrm{max}} & = \max_\vartheta v(\vartheta).
    \end{split}
\end{equation}
In the limit where we neglect curvature corrections, $v_\mathrm{min} = v_{\mathrm{max}}$ and is given exactly by the renormalized result given in Ref.~\cite{BistritzerMacDonald2011} for the continuum model.
Comparison of the velocity renormalization with and without the curvature corrections in this model is given in Fig.~\ref{fig:velocity_renormalization}.

\bibliography{mag}

\begin{thebibliography}{39}%
\makeatletter
\providecommand \@ifxundefined [1]{%
 \@ifx{#1\undefined}
}%
\providecommand \@ifnum [1]{%
 \ifnum #1\expandafter \@firstoftwo
 \else \expandafter \@secondoftwo
 \fi
}%
\providecommand \@ifx [1]{%
 \ifx #1\expandafter \@firstoftwo
 \else \expandafter \@secondoftwo
 \fi
}%
\providecommand \natexlab [1]{#1}%
\providecommand \enquote  [1]{``#1''}%
\providecommand \bibnamefont  [1]{#1}%
\providecommand \bibfnamefont [1]{#1}%
\providecommand \citenamefont [1]{#1}%
\providecommand \href@noop [0]{\@secondoftwo}%
\providecommand \href [0]{\begingroup \@sanitize@url \@href}%
\providecommand \@href[1]{\@@startlink{#1}\@@href}%
\providecommand \@@href[1]{\endgroup#1\@@endlink}%
\providecommand \@sanitize@url [0]{\catcode `\\12\catcode `\$12\catcode
  `\&12\catcode `\#12\catcode `\^12\catcode `\_12\catcode `\%12\relax}%
\providecommand \@@startlink[1]{}%
\providecommand \@@endlink[0]{}%
\providecommand \url  [0]{\begingroup\@sanitize@url \@url }%
\providecommand \@url [1]{\endgroup\@href {#1}{\urlprefix }}%
\providecommand \urlprefix  [0]{URL }%
\providecommand \Eprint [0]{\href }%
\providecommand \doibase [0]{http://dx.doi.org/}%
\providecommand \selectlanguage [0]{\@gobble}%
\providecommand \bibinfo  [0]{\@secondoftwo}%
\providecommand \bibfield  [0]{\@secondoftwo}%
\providecommand \translation [1]{[#1]}%
\providecommand \BibitemOpen [0]{}%
\providecommand \bibitemStop [0]{}%
\providecommand \bibitemNoStop [0]{.\EOS\space}%
\providecommand \EOS [0]{\spacefactor3000\relax}%
\providecommand \BibitemShut  [1]{\csname bibitem#1\endcsname}%
\let\auto@bib@innerbib\@empty
\bibitem [{\citenamefont {Geim}(2009)}]{Geim-2009}%
  \BibitemOpen
  \bibfield  {author} {\bibinfo {author} {\bibfnamefont {A.~K.}\ \bibnamefont
  {Geim}},\ }\href {\doibase 10.1126/science.1158877} {\bibfield  {journal}
  {\bibinfo  {journal} {Science}\ }\textbf {\bibinfo {volume} {324}},\ \bibinfo
  {pages} {1530} (\bibinfo {year} {2009})}\BibitemShut {NoStop}%
\bibitem [{\citenamefont {Geim}\ and\ \citenamefont
  {Grigorieva}(2013)}]{Geim-2013}%
  \BibitemOpen
  \bibfield  {author} {\bibinfo {author} {\bibfnamefont {A.~K.}\ \bibnamefont
  {Geim}}\ and\ \bibinfo {author} {\bibfnamefont {I.~V.}\ \bibnamefont
  {Grigorieva}},\ }\href {\doibase 10.1038/nature12385} {\bibfield  {journal}
  {\bibinfo  {journal} {Nature}\ }\textbf {\bibinfo {volume} {499}},\ \bibinfo
  {pages} {419} (\bibinfo {year} {2013})}\BibitemShut {NoStop}%
\bibitem [{\citenamefont {Li}\ \emph {et~al.}(2010{\natexlab{a}})\citenamefont
  {Li}, \citenamefont {Luican}, \citenamefont {Lopes~dos Santos}, \citenamefont
  {Castro~Neto}, \citenamefont {Reina}, \citenamefont {Kong},\ and\
  \citenamefont {Andrei}}]{li_observation_2010}%
  \BibitemOpen
  \bibfield  {author} {\bibinfo {author} {\bibfnamefont {G.}~\bibnamefont
  {Li}}, \bibinfo {author} {\bibfnamefont {A.}~\bibnamefont {Luican}}, \bibinfo
  {author} {\bibfnamefont {J.~M.~B.}\ \bibnamefont {Lopes~dos Santos}},
  \bibinfo {author} {\bibfnamefont {A.~H.}\ \bibnamefont {Castro~Neto}},
  \bibinfo {author} {\bibfnamefont {A.}~\bibnamefont {Reina}}, \bibinfo
  {author} {\bibfnamefont {J.}~\bibnamefont {Kong}}, \ and\ \bibinfo {author}
  {\bibfnamefont {E.~Y.}\ \bibnamefont {Andrei}},\ }\href {\doibase
  10.1038/nphys1463} {\bibfield  {journal} {\bibinfo  {journal} {Nat. Phys.}\
  }\textbf {\bibinfo {volume} {6}},\ \bibinfo {pages} {109} (\bibinfo {year}
  {2010}{\natexlab{a}})}\BibitemShut {NoStop}%
\bibitem [{\citenamefont {Luican}\ \emph {et~al.}(2011)\citenamefont {Luican},
  \citenamefont {Li}, \citenamefont {Reina}, \citenamefont {Kong},
  \citenamefont {Nair}, \citenamefont {Novoselov}, \citenamefont {Geim},\ and\
  \citenamefont {Andrei}}]{Luican-2011}%
  \BibitemOpen
  \bibfield  {author} {\bibinfo {author} {\bibfnamefont {A.}~\bibnamefont
  {Luican}}, \bibinfo {author} {\bibfnamefont {G.}~\bibnamefont {Li}}, \bibinfo
  {author} {\bibfnamefont {A.}~\bibnamefont {Reina}}, \bibinfo {author}
  {\bibfnamefont {J.}~\bibnamefont {Kong}}, \bibinfo {author} {\bibfnamefont
  {R.~R.}\ \bibnamefont {Nair}}, \bibinfo {author} {\bibfnamefont {K.~S.}\
  \bibnamefont {Novoselov}}, \bibinfo {author} {\bibfnamefont {A.~K.}\
  \bibnamefont {Geim}}, \ and\ \bibinfo {author} {\bibfnamefont {E.~Y.}\
  \bibnamefont {Andrei}},\ }\href {\doibase 10.1103/PhysRevLett.106.126802}
  {\bibfield  {journal} {\bibinfo  {journal} {Phys. Rev. Lett.}\ }\textbf
  {\bibinfo {volume} {106}},\ \bibinfo {pages} {126802} (\bibinfo {year}
  {2011})}\BibitemShut {NoStop}%
\bibitem [{\citenamefont {Cao}\ \emph {et~al.}(2018{\natexlab{a}})\citenamefont
  {Cao}, \citenamefont {Fatemi}, \citenamefont {Demir}, \citenamefont {Fang},
  \citenamefont {Tomarken}, \citenamefont {Luo}, \citenamefont
  {Sanchez-Yamagishi}, \citenamefont {Watanabe}, \citenamefont {Taniguchi},
  \citenamefont {Kaxiras} \emph {et~al.}}]{CaoJarillo2018a}%
  \BibitemOpen
  \bibfield  {author} {\bibinfo {author} {\bibfnamefont {Y.}~\bibnamefont
  {Cao}}, \bibinfo {author} {\bibfnamefont {V.}~\bibnamefont {Fatemi}},
  \bibinfo {author} {\bibfnamefont {A.}~\bibnamefont {Demir}}, \bibinfo
  {author} {\bibfnamefont {S.}~\bibnamefont {Fang}}, \bibinfo {author}
  {\bibfnamefont {S.~L.}\ \bibnamefont {Tomarken}}, \bibinfo {author}
  {\bibfnamefont {J.~Y.}\ \bibnamefont {Luo}}, \bibinfo {author} {\bibfnamefont
  {J.~D.}\ \bibnamefont {Sanchez-Yamagishi}}, \bibinfo {author} {\bibfnamefont
  {K.}~\bibnamefont {Watanabe}}, \bibinfo {author} {\bibfnamefont
  {T.}~\bibnamefont {Taniguchi}}, \bibinfo {author} {\bibfnamefont
  {E.}~\bibnamefont {Kaxiras}},  \emph {et~al.},\ }\href {\doibase
  10.1038/nature26154} {\bibfield  {journal} {\bibinfo  {journal} {Nature}\
  }\textbf {\bibinfo {volume} {556}},\ \bibinfo {pages} {80} (\bibinfo {year}
  {2018}{\natexlab{a}})}\BibitemShut {NoStop}%
\bibitem [{\citenamefont {Cao}\ \emph {et~al.}(2018{\natexlab{b}})\citenamefont
  {Cao}, \citenamefont {Fatemi}, \citenamefont {Fang}, \citenamefont
  {Watanabe}, \citenamefont {Taniguchi}, \citenamefont {Kaxiras},\ and\
  \citenamefont {Jarillo-Herrero}}]{CaoJarillo2018b}%
  \BibitemOpen
  \bibfield  {author} {\bibinfo {author} {\bibfnamefont {Y.}~\bibnamefont
  {Cao}}, \bibinfo {author} {\bibfnamefont {V.}~\bibnamefont {Fatemi}},
  \bibinfo {author} {\bibfnamefont {S.}~\bibnamefont {Fang}}, \bibinfo {author}
  {\bibfnamefont {K.}~\bibnamefont {Watanabe}}, \bibinfo {author}
  {\bibfnamefont {T.}~\bibnamefont {Taniguchi}}, \bibinfo {author}
  {\bibfnamefont {E.}~\bibnamefont {Kaxiras}}, \ and\ \bibinfo {author}
  {\bibfnamefont {P.}~\bibnamefont {Jarillo-Herrero}},\ }\href {\doibase
  10.1038/nature26160} {\bibfield  {journal} {\bibinfo  {journal} {Nature}\
  }\textbf {\bibinfo {volume} {556}},\ \bibinfo {pages} {43} (\bibinfo {year}
  {2018}{\natexlab{b}})}\BibitemShut {NoStop}%
\bibitem [{\citenamefont {Yankowitz}\ \emph {et~al.}(2019)\citenamefont
  {Yankowitz}, \citenamefont {Chen}, \citenamefont {Polshyn}, \citenamefont
  {Zhang}, \citenamefont {Watanabe}, \citenamefont {Taniguchi}, \citenamefont
  {Graf}, \citenamefont {Young},\ and\ \citenamefont {Dean}}]{Yankowitz-2019}%
  \BibitemOpen
  \bibfield  {author} {\bibinfo {author} {\bibfnamefont {M.}~\bibnamefont
  {Yankowitz}}, \bibinfo {author} {\bibfnamefont {S.}~\bibnamefont {Chen}},
  \bibinfo {author} {\bibfnamefont {H.}~\bibnamefont {Polshyn}}, \bibinfo
  {author} {\bibfnamefont {Y.}~\bibnamefont {Zhang}}, \bibinfo {author}
  {\bibfnamefont {K.}~\bibnamefont {Watanabe}}, \bibinfo {author}
  {\bibfnamefont {T.}~\bibnamefont {Taniguchi}}, \bibinfo {author}
  {\bibfnamefont {D.}~\bibnamefont {Graf}}, \bibinfo {author} {\bibfnamefont
  {A.~F.}\ \bibnamefont {Young}}, \ and\ \bibinfo {author} {\bibfnamefont
  {C.~R.}\ \bibnamefont {Dean}},\ }\href {\doibase 10.1126/science.aav1910}
  {\bibfield  {journal} {\bibinfo  {journal} {Science}\ }\textbf {\bibinfo
  {volume} {363}},\ \bibinfo {pages} {1059} (\bibinfo {year}
  {2019})}\BibitemShut {NoStop}%
\bibitem [{\citenamefont {Lu}\ \emph {et~al.}(2019)\citenamefont {Lu},
  \citenamefont {Stepanov}, \citenamefont {Yang}, \citenamefont {Xie},
  \citenamefont {Aamir}, \citenamefont {Das}, \citenamefont {Urgell},
  \citenamefont {Watanabe}, \citenamefont {Taniguchi}, \citenamefont {Zhang}
  \emph {et~al.}}]{Lu-2019}%
  \BibitemOpen
  \bibfield  {author} {\bibinfo {author} {\bibfnamefont {X.}~\bibnamefont
  {Lu}}, \bibinfo {author} {\bibfnamefont {P.}~\bibnamefont {Stepanov}},
  \bibinfo {author} {\bibfnamefont {W.}~\bibnamefont {Yang}}, \bibinfo {author}
  {\bibfnamefont {M.}~\bibnamefont {Xie}}, \bibinfo {author} {\bibfnamefont
  {M.~A.}\ \bibnamefont {Aamir}}, \bibinfo {author} {\bibfnamefont
  {I.}~\bibnamefont {Das}}, \bibinfo {author} {\bibfnamefont {C.}~\bibnamefont
  {Urgell}}, \bibinfo {author} {\bibfnamefont {K.}~\bibnamefont {Watanabe}},
  \bibinfo {author} {\bibfnamefont {T.}~\bibnamefont {Taniguchi}}, \bibinfo
  {author} {\bibfnamefont {G.}~\bibnamefont {Zhang}},  \emph {et~al.},\ }\href
  {https://arxiv.org/abs/1903.06513} {\bibfield  {journal} {\bibinfo  {journal}
  {arXiv preprint arXiv:1903.06513}\ } (\bibinfo {year} {2019})}\BibitemShut
  {NoStop}%
\bibitem [{\citenamefont {Das~Sarma}\ \emph {et~al.}(2011)\citenamefont
  {Das~Sarma}, \citenamefont {Adam}, \citenamefont {Hwang},\ and\ \citenamefont
  {Rossi}}]{DasSarma-2011}%
  \BibitemOpen
  \bibfield  {author} {\bibinfo {author} {\bibfnamefont {S.}~\bibnamefont
  {Das~Sarma}}, \bibinfo {author} {\bibfnamefont {S.}~\bibnamefont {Adam}},
  \bibinfo {author} {\bibfnamefont {E.~H.}\ \bibnamefont {Hwang}}, \ and\
  \bibinfo {author} {\bibfnamefont {E.}~\bibnamefont {Rossi}},\ }\href
  {\doibase 10.1103/RevModPhys.83.407} {\bibfield  {journal} {\bibinfo
  {journal} {Rev. Mod. Phys.}\ }\textbf {\bibinfo {volume} {83}},\ \bibinfo
  {pages} {407} (\bibinfo {year} {2011})}\BibitemShut {NoStop}%
\bibitem [{\citenamefont {Zibrov}\ \emph {et~al.}(2017)\citenamefont {Zibrov},
  \citenamefont {Kometter}, \citenamefont {Zhou}, \citenamefont {Spanton},
  \citenamefont {Taniguchi}, \citenamefont {Watanabe}, \citenamefont
  {Zaletel},\ and\ \citenamefont {Young}}]{Zibrov-2017}%
  \BibitemOpen
  \bibfield  {author} {\bibinfo {author} {\bibfnamefont {A.}~\bibnamefont
  {Zibrov}}, \bibinfo {author} {\bibfnamefont {C.}~\bibnamefont {Kometter}},
  \bibinfo {author} {\bibfnamefont {H.}~\bibnamefont {Zhou}}, \bibinfo {author}
  {\bibfnamefont {E.}~\bibnamefont {Spanton}}, \bibinfo {author} {\bibfnamefont
  {T.}~\bibnamefont {Taniguchi}}, \bibinfo {author} {\bibfnamefont
  {K.}~\bibnamefont {Watanabe}}, \bibinfo {author} {\bibfnamefont
  {M.}~\bibnamefont {Zaletel}}, \ and\ \bibinfo {author} {\bibfnamefont
  {A.}~\bibnamefont {Young}},\ }\href {\doibase 10.1038/nature23893} {\bibfield
   {journal} {\bibinfo  {journal} {Nature}\ }\textbf {\bibinfo {volume}
  {549}},\ \bibinfo {pages} {360} (\bibinfo {year} {2017})}\BibitemShut
  {NoStop}%
\bibitem [{\citenamefont {Du}\ \emph {et~al.}(2009)\citenamefont {Du},
  \citenamefont {Skachko}, \citenamefont {Duerr}, \citenamefont {Luican},\ and\
  \citenamefont {Andrei}}]{Du-2009}%
  \BibitemOpen
  \bibfield  {author} {\bibinfo {author} {\bibfnamefont {X.}~\bibnamefont
  {Du}}, \bibinfo {author} {\bibfnamefont {I.}~\bibnamefont {Skachko}},
  \bibinfo {author} {\bibfnamefont {F.}~\bibnamefont {Duerr}}, \bibinfo
  {author} {\bibfnamefont {A.}~\bibnamefont {Luican}}, \ and\ \bibinfo {author}
  {\bibfnamefont {E.~Y.}\ \bibnamefont {Andrei}},\ }\href {\doibase
  10.1038/nature08522} {\bibfield  {journal} {\bibinfo  {journal} {Nature}\
  }\textbf {\bibinfo {volume} {462}},\ \bibinfo {pages} {192} (\bibinfo {year}
  {2009})}\BibitemShut {NoStop}%
\bibitem [{\citenamefont {Kim}\ \emph {et~al.}(2016)\citenamefont {Kim},
  \citenamefont {Yankowitz}, \citenamefont {Fallahazad}, \citenamefont {Kang},
  \citenamefont {Movva}, \citenamefont {Huang}, \citenamefont {Larentis},
  \citenamefont {Corbet}, \citenamefont {Taniguchi}, \citenamefont {Watanabe}
  \emph {et~al.}}]{Kim-2016}%
  \BibitemOpen
  \bibfield  {author} {\bibinfo {author} {\bibfnamefont {K.}~\bibnamefont
  {Kim}}, \bibinfo {author} {\bibfnamefont {M.}~\bibnamefont {Yankowitz}},
  \bibinfo {author} {\bibfnamefont {B.}~\bibnamefont {Fallahazad}}, \bibinfo
  {author} {\bibfnamefont {S.}~\bibnamefont {Kang}}, \bibinfo {author}
  {\bibfnamefont {H.~C.}\ \bibnamefont {Movva}}, \bibinfo {author}
  {\bibfnamefont {S.}~\bibnamefont {Huang}}, \bibinfo {author} {\bibfnamefont
  {S.}~\bibnamefont {Larentis}}, \bibinfo {author} {\bibfnamefont {C.~M.}\
  \bibnamefont {Corbet}}, \bibinfo {author} {\bibfnamefont {T.}~\bibnamefont
  {Taniguchi}}, \bibinfo {author} {\bibfnamefont {K.}~\bibnamefont {Watanabe}},
   \emph {et~al.},\ }\href {\doibase 10.1021/acs.nanolett.5b05263} {\bibfield
  {journal} {\bibinfo  {journal} {Nano letters}\ }\textbf {\bibinfo {volume}
  {16}},\ \bibinfo {pages} {1989} (\bibinfo {year} {2016})}\BibitemShut
  {NoStop}%
\bibitem [{\citenamefont {Brihuega}\ \emph {et~al.}(2012)\citenamefont
  {Brihuega}, \citenamefont {Mallet}, \citenamefont {González-Herrero},
  \citenamefont {Trambly~de Laissardière}, \citenamefont {Ugeda},
  \citenamefont {Magaud}, \citenamefont {Gómez-Rodríguez}, \citenamefont
  {Ynduráin},\ and\ \citenamefont {Veuillen}}]{brihuega_unraveling_2012}%
  \BibitemOpen
  \bibfield  {author} {\bibinfo {author} {\bibfnamefont {I.}~\bibnamefont
  {Brihuega}}, \bibinfo {author} {\bibfnamefont {P.}~\bibnamefont {Mallet}},
  \bibinfo {author} {\bibfnamefont {H.}~\bibnamefont {González-Herrero}},
  \bibinfo {author} {\bibfnamefont {G.}~\bibnamefont {Trambly~de
  Laissardière}}, \bibinfo {author} {\bibfnamefont {M.~M.}\ \bibnamefont
  {Ugeda}}, \bibinfo {author} {\bibfnamefont {L.}~\bibnamefont {Magaud}},
  \bibinfo {author} {\bibfnamefont {J.~M.}\ \bibnamefont {Gómez-Rodríguez}},
  \bibinfo {author} {\bibfnamefont {F.}~\bibnamefont {Ynduráin}}, \ and\
  \bibinfo {author} {\bibfnamefont {J.-Y.}\ \bibnamefont {Veuillen}},\ }\href
  {\doibase 10.1103/PhysRevLett.109.196802} {\bibfield  {journal} {\bibinfo
  {journal} {Phys. Rev. Lett.}\ }\textbf {\bibinfo {volume} {109}},\ \bibinfo
  {pages} {196802} (\bibinfo {year} {2012})}\BibitemShut {NoStop}%
\bibitem [{\citenamefont {Wong}\ \emph {et~al.}(2015)\citenamefont {Wong},
  \citenamefont {Wang}, \citenamefont {Jung}, \citenamefont {Pezzini},
  \citenamefont {DaSilva}, \citenamefont {Tsai}, \citenamefont {Jung},
  \citenamefont {Khajeh}, \citenamefont {Kim}, \citenamefont {Lee},
  \citenamefont {Kahn}, \citenamefont {Tollabimazraehno}, \citenamefont
  {Rasool}, \citenamefont {Watanabe}, \citenamefont {Taniguchi}, \citenamefont
  {Zettl}, \citenamefont {Adam}, \citenamefont {MacDonald},\ and\ \citenamefont
  {Crommie}}]{wong_local_2015}%
  \BibitemOpen
  \bibfield  {author} {\bibinfo {author} {\bibfnamefont {D.}~\bibnamefont
  {Wong}}, \bibinfo {author} {\bibfnamefont {Y.}~\bibnamefont {Wang}}, \bibinfo
  {author} {\bibfnamefont {J.}~\bibnamefont {Jung}}, \bibinfo {author}
  {\bibfnamefont {S.}~\bibnamefont {Pezzini}}, \bibinfo {author} {\bibfnamefont
  {A.~M.}\ \bibnamefont {DaSilva}}, \bibinfo {author} {\bibfnamefont {H.-Z.}\
  \bibnamefont {Tsai}}, \bibinfo {author} {\bibfnamefont {H.~S.}\ \bibnamefont
  {Jung}}, \bibinfo {author} {\bibfnamefont {R.}~\bibnamefont {Khajeh}},
  \bibinfo {author} {\bibfnamefont {Y.}~\bibnamefont {Kim}}, \bibinfo {author}
  {\bibfnamefont {J.}~\bibnamefont {Lee}}, \bibinfo {author} {\bibfnamefont
  {S.}~\bibnamefont {Kahn}}, \bibinfo {author} {\bibfnamefont {S.}~\bibnamefont
  {Tollabimazraehno}}, \bibinfo {author} {\bibfnamefont {H.}~\bibnamefont
  {Rasool}}, \bibinfo {author} {\bibfnamefont {K.}~\bibnamefont {Watanabe}},
  \bibinfo {author} {\bibfnamefont {T.}~\bibnamefont {Taniguchi}}, \bibinfo
  {author} {\bibfnamefont {A.}~\bibnamefont {Zettl}}, \bibinfo {author}
  {\bibfnamefont {S.}~\bibnamefont {Adam}}, \bibinfo {author} {\bibfnamefont
  {A.~H.}\ \bibnamefont {MacDonald}}, \ and\ \bibinfo {author} {\bibfnamefont
  {M.~F.}\ \bibnamefont {Crommie}},\ }\href {\doibase
  10.1103/PhysRevB.92.155409} {\bibfield  {journal} {\bibinfo  {journal} {Phys.
  Rev. B}\ }\textbf {\bibinfo {volume} {92}},\ \bibinfo {pages} {155409}
  (\bibinfo {year} {2015})}\BibitemShut {NoStop}%
\bibitem [{\citenamefont {Kerelsky}\ \emph {et~al.}(2018)\citenamefont
  {Kerelsky}, \citenamefont {McGilly}, \citenamefont {Kennes}, \citenamefont
  {Xian}, \citenamefont {Yankowitz}, \citenamefont {Chen}, \citenamefont
  {Watanabe}, \citenamefont {Taniguchi}, \citenamefont {Hone}, \citenamefont
  {Dean}, \citenamefont {Rubio},\ and\ \citenamefont
  {Pasupathy}}]{kerelsky_magic_2018}%
  \BibitemOpen
  \bibfield  {author} {\bibinfo {author} {\bibfnamefont {A.}~\bibnamefont
  {Kerelsky}}, \bibinfo {author} {\bibfnamefont {L.}~\bibnamefont {McGilly}},
  \bibinfo {author} {\bibfnamefont {D.~M.}\ \bibnamefont {Kennes}}, \bibinfo
  {author} {\bibfnamefont {L.}~\bibnamefont {Xian}}, \bibinfo {author}
  {\bibfnamefont {M.}~\bibnamefont {Yankowitz}}, \bibinfo {author}
  {\bibfnamefont {S.}~\bibnamefont {Chen}}, \bibinfo {author} {\bibfnamefont
  {K.}~\bibnamefont {Watanabe}}, \bibinfo {author} {\bibfnamefont
  {T.}~\bibnamefont {Taniguchi}}, \bibinfo {author} {\bibfnamefont
  {J.}~\bibnamefont {Hone}}, \bibinfo {author} {\bibfnamefont {C.}~\bibnamefont
  {Dean}}, \bibinfo {author} {\bibfnamefont {A.}~\bibnamefont {Rubio}}, \ and\
  \bibinfo {author} {\bibfnamefont {A.~N.}\ \bibnamefont {Pasupathy}},\ }\href
  {http://arxiv.org/abs/1812.08776} {\bibfield  {journal} {\bibinfo  {journal}
  {arXiv:1812.08776 [cond-mat]}\ } (\bibinfo {year} {2018})},\ \bibinfo {note}
  {arXiv: 1812.08776}\BibitemShut {NoStop}%
\bibitem [{\citenamefont {Choi}\ \emph {et~al.}(2019)\citenamefont {Choi},
  \citenamefont {Kemmer}, \citenamefont {Peng}, \citenamefont {Thomson},
  \citenamefont {Arora}, \citenamefont {Polski}, \citenamefont {Zhang},
  \citenamefont {Ren}, \citenamefont {Alicea}, \citenamefont {Refael},
  \citenamefont {von Oppen}, \citenamefont {Watanabe}, \citenamefont
  {Taniguchi},\ and\ \citenamefont {Nadj-Perge}}]{choi_imaging_2019}%
  \BibitemOpen
  \bibfield  {author} {\bibinfo {author} {\bibfnamefont {Y.}~\bibnamefont
  {Choi}}, \bibinfo {author} {\bibfnamefont {J.}~\bibnamefont {Kemmer}},
  \bibinfo {author} {\bibfnamefont {Y.}~\bibnamefont {Peng}}, \bibinfo {author}
  {\bibfnamefont {A.}~\bibnamefont {Thomson}}, \bibinfo {author} {\bibfnamefont
  {H.}~\bibnamefont {Arora}}, \bibinfo {author} {\bibfnamefont
  {R.}~\bibnamefont {Polski}}, \bibinfo {author} {\bibfnamefont
  {Y.}~\bibnamefont {Zhang}}, \bibinfo {author} {\bibfnamefont
  {H.}~\bibnamefont {Ren}}, \bibinfo {author} {\bibfnamefont {J.}~\bibnamefont
  {Alicea}}, \bibinfo {author} {\bibfnamefont {G.}~\bibnamefont {Refael}},
  \bibinfo {author} {\bibfnamefont {F.}~\bibnamefont {von Oppen}}, \bibinfo
  {author} {\bibfnamefont {K.}~\bibnamefont {Watanabe}}, \bibinfo {author}
  {\bibfnamefont {T.}~\bibnamefont {Taniguchi}}, \ and\ \bibinfo {author}
  {\bibfnamefont {S.}~\bibnamefont {Nadj-Perge}},\ }\href
  {http://arxiv.org/abs/1901.02997} {\bibfield  {journal} {\bibinfo  {journal}
  {arXiv:1901.02997 [cond-mat]}\ } (\bibinfo {year} {2019})},\ \bibinfo {note}
  {arXiv: 1901.02997}\BibitemShut {NoStop}%
\bibitem [{\citenamefont {Jiang}\ \emph {et~al.}(2019)\citenamefont {Jiang},
  \citenamefont {Lai}, \citenamefont {Watanabe}, \citenamefont {Taniguchi},
  \citenamefont {Haule}, \citenamefont {Mao},\ and\ \citenamefont
  {Andrei}}]{jiang_charge-order_2019}%
  \BibitemOpen
  \bibfield  {author} {\bibinfo {author} {\bibfnamefont {Y.}~\bibnamefont
  {Jiang}}, \bibinfo {author} {\bibfnamefont {X.}~\bibnamefont {Lai}}, \bibinfo
  {author} {\bibfnamefont {K.}~\bibnamefont {Watanabe}}, \bibinfo {author}
  {\bibfnamefont {T.}~\bibnamefont {Taniguchi}}, \bibinfo {author}
  {\bibfnamefont {K.}~\bibnamefont {Haule}}, \bibinfo {author} {\bibfnamefont
  {J.}~\bibnamefont {Mao}}, \ and\ \bibinfo {author} {\bibfnamefont {E.~Y.}\
  \bibnamefont {Andrei}},\ }\href {\doibase 10.1038/s41586-019-1460-4}
  {\bibfield  {journal} {\bibinfo  {journal} {Nature}\ ,\ \bibinfo {pages} {1}}
  (\bibinfo {year} {2019})}\BibitemShut {NoStop}%
\bibitem [{\citenamefont {Xie}\ \emph {et~al.}(2019)\citenamefont {Xie},
  \citenamefont {Lian}, \citenamefont {Jäck}, \citenamefont {Liu},
  \citenamefont {Chiu}, \citenamefont {Watanabe}, \citenamefont {Taniguchi},
  \citenamefont {Bernevig},\ and\ \citenamefont
  {Yazdani}}]{xie_spectroscopic_2019}%
  \BibitemOpen
  \bibfield  {author} {\bibinfo {author} {\bibfnamefont {Y.}~\bibnamefont
  {Xie}}, \bibinfo {author} {\bibfnamefont {B.}~\bibnamefont {Lian}}, \bibinfo
  {author} {\bibfnamefont {B.}~\bibnamefont {Jäck}}, \bibinfo {author}
  {\bibfnamefont {X.}~\bibnamefont {Liu}}, \bibinfo {author} {\bibfnamefont
  {C.-L.}\ \bibnamefont {Chiu}}, \bibinfo {author} {\bibfnamefont
  {K.}~\bibnamefont {Watanabe}}, \bibinfo {author} {\bibfnamefont
  {T.}~\bibnamefont {Taniguchi}}, \bibinfo {author} {\bibfnamefont {B.~A.}\
  \bibnamefont {Bernevig}}, \ and\ \bibinfo {author} {\bibfnamefont
  {A.}~\bibnamefont {Yazdani}},\ }\href {\doibase 10.1038/s41586-019-1422-x}
  {\bibfield  {journal} {\bibinfo  {journal} {Nature}\ }\textbf {\bibinfo
  {volume} {572}},\ \bibinfo {pages} {101} (\bibinfo {year}
  {2019})}\BibitemShut {NoStop}%
\bibitem [{\citenamefont {Andrei}(2019)}]{Priv_Comm_Eva}%
  \BibitemOpen
  \bibfield  {author} {\bibinfo {author} {\bibfnamefont {E.~Y.}\ \bibnamefont
  {Andrei}},\ }\href@noop {} {}\bibinfo {howpublished} {{Private
  Communication}} (\bibinfo {year} {2019})\BibitemShut {NoStop}%
\bibitem [{\citenamefont {Lopes~dos Santos}\ \emph {et~al.}(2007)\citenamefont
  {Lopes~dos Santos}, \citenamefont {Peres},\ and\ \citenamefont
  {Castro~Neto}}]{lopesdossantos_graphene_2007}%
  \BibitemOpen
  \bibfield  {author} {\bibinfo {author} {\bibfnamefont {J.~M.~B.}\
  \bibnamefont {Lopes~dos Santos}}, \bibinfo {author} {\bibfnamefont
  {N.~M.~R.}\ \bibnamefont {Peres}}, \ and\ \bibinfo {author} {\bibfnamefont
  {A.~H.}\ \bibnamefont {Castro~Neto}},\ }\href {\doibase
  10.1103/PhysRevLett.99.256802} {\bibfield  {journal} {\bibinfo  {journal}
  {Phys. Rev. Lett.}\ }\textbf {\bibinfo {volume} {99}},\ \bibinfo {pages}
  {256802} (\bibinfo {year} {2007})}\BibitemShut {NoStop}%
\bibitem [{\citenamefont {Bistritzer}\ and\ \citenamefont
  {MacDonald}(2011)}]{BistritzerMacDonald2011}%
  \BibitemOpen
  \bibfield  {author} {\bibinfo {author} {\bibfnamefont {R.}~\bibnamefont
  {Bistritzer}}\ and\ \bibinfo {author} {\bibfnamefont {A.~H.}\ \bibnamefont
  {MacDonald}},\ }\href@noop {} {\bibfield  {journal} {\bibinfo  {journal}
  {Proc. Natl. Acad. Sci.}\ }\textbf {\bibinfo {volume} {108}},\ \bibinfo
  {pages} {12233} (\bibinfo {year} {2011})}\BibitemShut {NoStop}%
\bibitem [{\citenamefont {Mele}(2010)}]{mele_commensuration_2010}%
  \BibitemOpen
  \bibfield  {author} {\bibinfo {author} {\bibfnamefont {E.~J.}\ \bibnamefont
  {Mele}},\ }\href {\doibase 10.1103/PhysRevB.81.161405} {\bibfield  {journal}
  {\bibinfo  {journal} {Phys. Rev. B}\ }\textbf {\bibinfo {volume} {81}},\
  \bibinfo {pages} {161405(R)} (\bibinfo {year} {2010})}\BibitemShut {NoStop}%
\bibitem [{\citenamefont {Moon}\ and\ \citenamefont
  {Koshino}(2012)}]{moon_energy_2012}%
  \BibitemOpen
  \bibfield  {author} {\bibinfo {author} {\bibfnamefont {P.}~\bibnamefont
  {Moon}}\ and\ \bibinfo {author} {\bibfnamefont {M.}~\bibnamefont {Koshino}},\
  }\href {\doibase 10.1103/PhysRevB.85.195458} {\bibfield  {journal} {\bibinfo
  {journal} {Phys. Rev. B}\ }\textbf {\bibinfo {volume} {85}},\ \bibinfo
  {pages} {195458} (\bibinfo {year} {2012})}\BibitemShut {NoStop}%
\bibitem [{\citenamefont {Fu}\ \emph {et~al.}(2018)\citenamefont {Fu},
  \citenamefont {K{\"o}nig}, \citenamefont {Wilson}, \citenamefont {Chou},\
  and\ \citenamefont {Pixley}}]{Fu-2018}%
  \BibitemOpen
  \bibfield  {author} {\bibinfo {author} {\bibfnamefont {Y.}~\bibnamefont
  {Fu}}, \bibinfo {author} {\bibfnamefont {E.}~\bibnamefont {K{\"o}nig}},
  \bibinfo {author} {\bibfnamefont {J.}~\bibnamefont {Wilson}}, \bibinfo
  {author} {\bibfnamefont {Y.-Z.}\ \bibnamefont {Chou}}, \ and\ \bibinfo
  {author} {\bibfnamefont {J.}~\bibnamefont {Pixley}},\ }\href
  {https://arxiv.org/abs/1809.04604} {\bibfield  {journal} {\bibinfo  {journal}
  {arXiv preprint arXiv:1809.04604}\ } (\bibinfo {year} {2018})}\BibitemShut
  {NoStop}%
\bibitem [{\citenamefont {Bi}\ \emph {et~al.}(2019)\citenamefont {Bi},
  \citenamefont {Yuan},\ and\ \citenamefont {Fu}}]{bi_designing_2019}%
  \BibitemOpen
  \bibfield  {author} {\bibinfo {author} {\bibfnamefont {Z.}~\bibnamefont
  {Bi}}, \bibinfo {author} {\bibfnamefont {N.~F.~Q.}\ \bibnamefont {Yuan}}, \
  and\ \bibinfo {author} {\bibfnamefont {L.}~\bibnamefont {Fu}},\ }\href
  {http://arxiv.org/abs/1902.10146} {\bibfield  {journal} {\bibinfo  {journal}
  {arXiv:1902.10146 [cond-mat]}\ } (\bibinfo {year} {2019})},\ \bibinfo {note}
  {arXiv: 1902.10146}\BibitemShut {NoStop}%
\bibitem [{\citenamefont {Wei\ss{}e}\ \emph {et~al.}(2006)\citenamefont
  {Wei\ss{}e}, \citenamefont {Wellein}, \citenamefont {Alvermann},\ and\
  \citenamefont {Fehske}}]{Weisse-2006}%
  \BibitemOpen
  \bibfield  {author} {\bibinfo {author} {\bibfnamefont {A.}~\bibnamefont
  {Wei\ss{}e}}, \bibinfo {author} {\bibfnamefont {G.}~\bibnamefont {Wellein}},
  \bibinfo {author} {\bibfnamefont {A.}~\bibnamefont {Alvermann}}, \ and\
  \bibinfo {author} {\bibfnamefont {H.}~\bibnamefont {Fehske}},\ }\href
  {\doibase 10.1103/RevModPhys.78.275} {\bibfield  {journal} {\bibinfo
  {journal} {Rev. Mod. Phys.}\ }\textbf {\bibinfo {volume} {78}},\ \bibinfo
  {pages} {275} (\bibinfo {year} {2006})}\BibitemShut {NoStop}%
\bibitem [{\citenamefont {Nam}\ and\ \citenamefont {Koshino}(2017)}]{Nam-2017}%
  \BibitemOpen
  \bibfield  {author} {\bibinfo {author} {\bibfnamefont {N.~N.~T.}\
  \bibnamefont {Nam}}\ and\ \bibinfo {author} {\bibfnamefont {M.}~\bibnamefont
  {Koshino}},\ }\href {\doibase 10.1103/PhysRevB.96.075311} {\bibfield
  {journal} {\bibinfo  {journal} {Phys. Rev. B}\ }\textbf {\bibinfo {volume}
  {96}},\ \bibinfo {pages} {075311} (\bibinfo {year} {2017})}\BibitemShut
  {NoStop}%
\bibitem [{\citenamefont {Carr}\ \emph {et~al.}(2019)\citenamefont {Carr},
  \citenamefont {Fang}, \citenamefont {Zhu},\ and\ \citenamefont
  {Kaxiras}}]{Carr-2019}%
  \BibitemOpen
  \bibfield  {author} {\bibinfo {author} {\bibfnamefont {S.}~\bibnamefont
  {Carr}}, \bibinfo {author} {\bibfnamefont {S.}~\bibnamefont {Fang}}, \bibinfo
  {author} {\bibfnamefont {Z.}~\bibnamefont {Zhu}}, \ and\ \bibinfo {author}
  {\bibfnamefont {E.}~\bibnamefont {Kaxiras}},\ }\href {\doibase
  10.1103/PhysRevResearch.1.013001} {\bibfield  {journal} {\bibinfo  {journal}
  {Phys. Rev. Research}\ }\textbf {\bibinfo {volume} {1}},\ \bibinfo {pages}
  {013001} (\bibinfo {year} {2019})}\BibitemShut {NoStop}%
\bibitem [{\citenamefont {Li}\ \emph {et~al.}(2010{\natexlab{b}})\citenamefont
  {Li}, \citenamefont {Luican}, \citenamefont {Lopes~dos Santos}, \citenamefont
  {Castro~Neto}, \citenamefont {Reina}, \citenamefont {Kong},\ and\
  \citenamefont {Andrei}}]{Li-2010}%
  \BibitemOpen
  \bibfield  {author} {\bibinfo {author} {\bibfnamefont {G.}~\bibnamefont
  {Li}}, \bibinfo {author} {\bibfnamefont {A.}~\bibnamefont {Luican}}, \bibinfo
  {author} {\bibfnamefont {J.}~\bibnamefont {Lopes~dos Santos}}, \bibinfo
  {author} {\bibfnamefont {A.}~\bibnamefont {Castro~Neto}}, \bibinfo {author}
  {\bibfnamefont {A.}~\bibnamefont {Reina}}, \bibinfo {author} {\bibfnamefont
  {J.}~\bibnamefont {Kong}}, \ and\ \bibinfo {author} {\bibfnamefont
  {E.}~\bibnamefont {Andrei}},\ }\href {\doibase 10.1038/nphys1463} {\bibfield
  {journal} {\bibinfo  {journal} {Nat. Phys.}\ }\textbf {\bibinfo {volume}
  {6}},\ \bibinfo {pages} {109} (\bibinfo {year}
  {2010}{\natexlab{b}})}\BibitemShut {NoStop}%
\bibitem [{\citenamefont {Lopes~dos Santos}\ \emph {et~al.}(2012)\citenamefont
  {Lopes~dos Santos}, \citenamefont {Peres},\ and\ \citenamefont
  {Castro~Neto}}]{DosSantosNeto2012}%
  \BibitemOpen
  \bibfield  {author} {\bibinfo {author} {\bibfnamefont {J.~M.~B.}\
  \bibnamefont {Lopes~dos Santos}}, \bibinfo {author} {\bibfnamefont
  {N.~M.~R.}\ \bibnamefont {Peres}}, \ and\ \bibinfo {author} {\bibfnamefont
  {A.~H.}\ \bibnamefont {Castro~Neto}},\ }\href {\doibase
  10.1103/PhysRevB.86.155449} {\bibfield  {journal} {\bibinfo  {journal} {Phys.
  Rev. B}\ }\textbf {\bibinfo {volume} {86}},\ \bibinfo {pages} {155449}
  (\bibinfo {year} {2012})}\BibitemShut {NoStop}%
\bibitem [{\citenamefont {Pixley}\ \emph {et~al.}(2018)\citenamefont {Pixley},
  \citenamefont {Wilson}, \citenamefont {Huse},\ and\ \citenamefont
  {Gopalakrishnan}}]{PixleyGopalakrishnan2018}%
  \BibitemOpen
  \bibfield  {author} {\bibinfo {author} {\bibfnamefont {J.~H.}\ \bibnamefont
  {Pixley}}, \bibinfo {author} {\bibfnamefont {J.~H.}\ \bibnamefont {Wilson}},
  \bibinfo {author} {\bibfnamefont {D.~A.}\ \bibnamefont {Huse}}, \ and\
  \bibinfo {author} {\bibfnamefont {S.}~\bibnamefont {Gopalakrishnan}},\ }\href
  {\doibase 10.1103/PhysRevLett.120.207604} {\bibfield  {journal} {\bibinfo
  {journal} {Phys. Rev. Lett.}\ }\textbf {\bibinfo {volume} {120}},\ \bibinfo
  {pages} {207604} (\bibinfo {year} {2018})}\BibitemShut {NoStop}%
\bibitem [{\citenamefont {Yao}\ \emph {et~al.}(2018{\natexlab{a}})\citenamefont
  {Yao}, \citenamefont {Wang}, \citenamefont {Bao}, \citenamefont {Zhang},
  \citenamefont {Zhang}, \citenamefont {Bao}, \citenamefont {Chan},
  \citenamefont {Chen}, \citenamefont {Avila}, \citenamefont {Asensio} \emph
  {et~al.}}]{Yao-2018}%
  \BibitemOpen
  \bibfield  {author} {\bibinfo {author} {\bibfnamefont {W.}~\bibnamefont
  {Yao}}, \bibinfo {author} {\bibfnamefont {E.}~\bibnamefont {Wang}}, \bibinfo
  {author} {\bibfnamefont {C.}~\bibnamefont {Bao}}, \bibinfo {author}
  {\bibfnamefont {Y.}~\bibnamefont {Zhang}}, \bibinfo {author} {\bibfnamefont
  {K.}~\bibnamefont {Zhang}}, \bibinfo {author} {\bibfnamefont
  {K.}~\bibnamefont {Bao}}, \bibinfo {author} {\bibfnamefont {C.~K.}\
  \bibnamefont {Chan}}, \bibinfo {author} {\bibfnamefont {C.}~\bibnamefont
  {Chen}}, \bibinfo {author} {\bibfnamefont {J.}~\bibnamefont {Avila}},
  \bibinfo {author} {\bibfnamefont {M.~C.}\ \bibnamefont {Asensio}},  \emph
  {et~al.},\ }\href {\doibase doi.org/10.1073/pnas.1720865115} {\bibfield
  {journal} {\bibinfo  {journal} {Proc. Natl. Acad. Sci.}\ }\textbf {\bibinfo
  {volume} {115}},\ \bibinfo {pages} {6928} (\bibinfo {year}
  {2018}{\natexlab{a}})}\BibitemShut {NoStop}%
\bibitem [{\citenamefont {Ahn}\ \emph {et~al.}(2018)\citenamefont {Ahn},
  \citenamefont {Moon}, \citenamefont {Kim}, \citenamefont {Kim}, \citenamefont
  {Shin}, \citenamefont {Kim}, \citenamefont {Cha}, \citenamefont {Kahng},
  \citenamefont {Kim}, \citenamefont {Koshino} \emph {et~al.}}]{Ahn-2018}%
  \BibitemOpen
  \bibfield  {author} {\bibinfo {author} {\bibfnamefont {S.~J.}\ \bibnamefont
  {Ahn}}, \bibinfo {author} {\bibfnamefont {P.}~\bibnamefont {Moon}}, \bibinfo
  {author} {\bibfnamefont {T.-H.}\ \bibnamefont {Kim}}, \bibinfo {author}
  {\bibfnamefont {H.-W.}\ \bibnamefont {Kim}}, \bibinfo {author} {\bibfnamefont
  {H.-C.}\ \bibnamefont {Shin}}, \bibinfo {author} {\bibfnamefont {E.~H.}\
  \bibnamefont {Kim}}, \bibinfo {author} {\bibfnamefont {H.~W.}\ \bibnamefont
  {Cha}}, \bibinfo {author} {\bibfnamefont {S.-J.}\ \bibnamefont {Kahng}},
  \bibinfo {author} {\bibfnamefont {P.}~\bibnamefont {Kim}}, \bibinfo {author}
  {\bibfnamefont {M.}~\bibnamefont {Koshino}},  \emph {et~al.},\ }\href
  {\doibase 10.1126/science.aar8412} {\bibfield  {journal} {\bibinfo  {journal}
  {Science}\ }\textbf {\bibinfo {volume} {361}},\ \bibinfo {pages} {782}
  (\bibinfo {year} {2018})}\BibitemShut {NoStop}%
\bibitem [{\citenamefont {Castro~Neto}\ \emph {et~al.}(2009)\citenamefont
  {Castro~Neto}, \citenamefont {Guinea}, \citenamefont {Peres}, \citenamefont
  {Novoselov},\ and\ \citenamefont {Geim}}]{castro_neto_electronic_2009}%
  \BibitemOpen
  \bibfield  {author} {\bibinfo {author} {\bibfnamefont {A.~H.}\ \bibnamefont
  {Castro~Neto}}, \bibinfo {author} {\bibfnamefont {F.}~\bibnamefont {Guinea}},
  \bibinfo {author} {\bibfnamefont {N.~M.~R.}\ \bibnamefont {Peres}}, \bibinfo
  {author} {\bibfnamefont {K.~S.}\ \bibnamefont {Novoselov}}, \ and\ \bibinfo
  {author} {\bibfnamefont {A.~K.}\ \bibnamefont {Geim}},\ }\href {\doibase
  10.1103/RevModPhys.81.109} {\bibfield  {journal} {\bibinfo  {journal} {Rev.
  Mod. Phys.}\ }\textbf {\bibinfo {volume} {81}},\ \bibinfo {pages} {109}
  (\bibinfo {year} {2009})}\BibitemShut {NoStop}%
\bibitem [{Note1()}]{Note1}%
  \BibitemOpen
  \bibinfo {note} {The phases always modify terms like $\protect \mathbf q_j
  \cdot \protect \mathbf r + \phi _j$, and in order to see how they represent a
  center of rotation, consider an $\protect \mathbf q_j \cdot ( \protect
  \mathbf r - \protect \mathbf r_0)$, then $\phi _j = - \protect \mathbf q_j
  \cdot \protect \mathbf r $ and $\DOTSB \sum@ \slimits@ _j \phi _j = 0$. In
  fact, for any $\phi _{1,2}$ we can determine an $\protect \mathbf r_0$ that
  creates it.}\BibitemShut {Stop}%
\bibitem [{\citenamefont {Lin}\ and\ \citenamefont
  {Tom\'anek}(2018)}]{Lin-2018}%
  \BibitemOpen
  \bibfield  {author} {\bibinfo {author} {\bibfnamefont {X.}~\bibnamefont
  {Lin}}\ and\ \bibinfo {author} {\bibfnamefont {D.}~\bibnamefont
  {Tom\'anek}},\ }\href {\doibase 10.1103/PhysRevB.98.081410} {\bibfield
  {journal} {\bibinfo  {journal} {Phys. Rev. B}\ }\textbf {\bibinfo {volume}
  {98}},\ \bibinfo {pages} {081410} (\bibinfo {year} {2018})}\BibitemShut
  {NoStop}%
\bibitem [{\citenamefont {Tarnopolsky}\ \emph {et~al.}(2019)\citenamefont
  {Tarnopolsky}, \citenamefont {Kruchkov},\ and\ \citenamefont
  {Vishwanath}}]{Tarnopolsky-2019}%
  \BibitemOpen
  \bibfield  {author} {\bibinfo {author} {\bibfnamefont {G.}~\bibnamefont
  {Tarnopolsky}}, \bibinfo {author} {\bibfnamefont {A.~J.}\ \bibnamefont
  {Kruchkov}}, \ and\ \bibinfo {author} {\bibfnamefont {A.}~\bibnamefont
  {Vishwanath}},\ }\href {\doibase 10.1103/PhysRevLett.122.106405} {\bibfield
  {journal} {\bibinfo  {journal} {Phys. Rev. Lett.}\ }\textbf {\bibinfo
  {volume} {122}},\ \bibinfo {pages} {106405} (\bibinfo {year}
  {2019})}\BibitemShut {NoStop}%
\bibitem [{\citenamefont {Lin}\ \emph {et~al.}(2018)\citenamefont {Lin},
  \citenamefont {Liu},\ and\ \citenamefont {Tom\'anek}}]{Lin2-2018}%
  \BibitemOpen
  \bibfield  {author} {\bibinfo {author} {\bibfnamefont {X.}~\bibnamefont
  {Lin}}, \bibinfo {author} {\bibfnamefont {D.}~\bibnamefont {Liu}}, \ and\
  \bibinfo {author} {\bibfnamefont {D.}~\bibnamefont {Tom\'anek}},\ }\href
  {\doibase 10.1103/PhysRevB.98.195432} {\bibfield  {journal} {\bibinfo
  {journal} {Phys. Rev. B}\ }\textbf {\bibinfo {volume} {98}},\ \bibinfo
  {pages} {195432} (\bibinfo {year} {2018})}\BibitemShut {NoStop}%
\bibitem [{\citenamefont {Yao}\ \emph {et~al.}(2018{\natexlab{b}})\citenamefont
  {Yao}, \citenamefont {Wang}, \citenamefont {Bao}, \citenamefont {Zhang},
  \citenamefont {Zhang}, \citenamefont {Bao}, \citenamefont {Chan},
  \citenamefont {Chen}, \citenamefont {Avila}, \citenamefont {Asensio} \emph
  {et~al.}}]{Yao2018}%
  \BibitemOpen
  \bibfield  {author} {\bibinfo {author} {\bibfnamefont {W.}~\bibnamefont
  {Yao}}, \bibinfo {author} {\bibfnamefont {E.}~\bibnamefont {Wang}}, \bibinfo
  {author} {\bibfnamefont {C.}~\bibnamefont {Bao}}, \bibinfo {author}
  {\bibfnamefont {Y.}~\bibnamefont {Zhang}}, \bibinfo {author} {\bibfnamefont
  {K.}~\bibnamefont {Zhang}}, \bibinfo {author} {\bibfnamefont
  {K.}~\bibnamefont {Bao}}, \bibinfo {author} {\bibfnamefont {C.~K.}\
  \bibnamefont {Chan}}, \bibinfo {author} {\bibfnamefont {C.}~\bibnamefont
  {Chen}}, \bibinfo {author} {\bibfnamefont {J.}~\bibnamefont {Avila}},
  \bibinfo {author} {\bibfnamefont {M.~C.}\ \bibnamefont {Asensio}},  \emph
  {et~al.},\ }\href {\doibase doi.org/10.1073/pnas.1720865115} {\bibfield
  {journal} {\bibinfo  {journal} {Proc. Natl. Acad. Sci.}\ }\textbf {\bibinfo
  {volume} {115}},\ \bibinfo {pages} {6928} (\bibinfo {year}
  {2018}{\natexlab{b}})}\BibitemShut {NoStop}%
\end{thebibliography}%

\end{document}